\documentclass[graybox]{svmult}

\usepackage{mathptmx}       
\usepackage{helvet}         
\usepackage{courier}        
\usepackage{type1cm}        
\usepackage{makeidx}         
\usepackage{graphicx}        
\usepackage{multicol}        
\usepackage[bottom]{footmisc}

\usepackage{siunitx}

\makeindex             

\begin{document}

\title*{Microwave studies of the fractional Josephson effect in HgTe-based Josephson junctions}
\titlerunning{Fractional Josephson effect in HgTe-based Josephson junctions}
\author{E. Bocquillon, J. Wiedenmann, R. S. Deacon, T. M. Klapwijk, H. Buhmann, L. W. Molenkamp}
\authorrunning{E. Bocquillon et al.}

\institute{E. Bocquillon \at Laboratoire Pierre Aigrain, Ecole Normale Sup\'erieure-PSL Research University, CNRS, Universit\'e Pierre et Marie Curie-Sorbonne Universit\'es, Universit\'e Paris Diderot-Sorbonne Paris Cit\'e,
24 rue Lhomond, 75231 Paris Cedex 05, France \\\email{erwann.bocquillon@lpa.ens.fr}
\and J. Wiedenmann, H. Buhmann, L.W. Molenkamp \at Physikalisches Institut (EP3), University of W\"urzburg, Am Hubland 97074 W\"urzburg (Germany)
\& Institute for Topological Insulators, Am Hubland, D-97074, Würzburg (Germany)
\and R.S. Deacon \at Center for Emergent Matter Science, RIKEN, 2-1 Hirosawa, Wako-shi, Saitama, 351-0198 (Japan) \& Advanced Device Laboratory, RIKEN, 2-1 Hirosawa, Wako-shi, Saitama, 351-0198 (Japan)
\and T. M. Klapwijk \at Kavli Institute of Nanoscience, Faculty of Applied Sciences, Delft University of Technology
Lorentzweg 1, 2628 CJ Delft (The Netherlands)
}

%
%
\maketitle

\abstract*{The rise of topological phases of matter is strongly connected to their potential to host Majorana bound states, a powerful ingredient in the search for a robust, topologically protected, quantum information processing. In order to produce such states, a method of choice is to induce superconductivity in topological insulators. The engineering of the interplay between superconductivity and the electronic properties of a topological insulator is a challenging task and it is consequently very important to understand the physics of simple superconducting devices such as Josephson junctions, in which new topological properties are expected to emerge. \newline\indent In this article, we review recent experiments investigating topological superconductivity in topological insulators, using microwave excitation and detection techniques. More precisely, we have fabricated and studied topological Josephson junctions made of HgTe weak links in contact with two Al or Nb contacts. In such devices, we have observed two signatures of the fractional Josephson effect, which is expected to emerge from topologically-protected gapless Andreev bound states. 
\newline\indent We first recall the theoretical background on topological Josephson junctions, then move to the experimental observations. Then, we assess the topological origin of the observed features and conclude with an outlook towards more advanced microwave spectroscopy experiments, currently under development.}

\abstract{The rise of topological phases of matter is strongly connected to their potential to host Majorana bound states, a powerful ingredient in the search for a robust, topologically protected, quantum information processing. In order to produce such states, a method of choice is to induce superconductivity in topological insulators. The engineering of the interplay between superconductivity and the electronic properties of a topological insulator is a challenging task and it is consequently very important to understand the physics of simple superconducting devices such as Josephson junctions, in which new topological properties are expected to emerge. \newline\indent In this article, we review recent experiments investigating topological superconductivity in topological insulators, using microwave excitation and detection techniques. More precisely, we have fabricated and studied topological Josephson junctions made of HgTe weak links in contact with Al or Nb contacts. In such devices, we have observed two signatures of the fractional Josephson effect, which is expected to emerge from topologically-protected gapless Andreev bound states. 
\newline\indent We first recall the theoretical background on topological Josephson junctions, then move to the experimental observations. Then, we assess the topological origin of the observed features and conclude with an outlook towards more advanced microwave spectroscopy experiments, currently under development.}
%


\section{Gapless Andreev bound states in topological Josephson junctions}
\label{sec:1}

\begin{svgraybox}
In this first section, we recall the basic ingredients of induced superconductivity in topological insulators. The broken spin rotation symmetry in these systems results in the formation of a peculiar phase with a $p$-wave symmetry. We briefly introduce an important consequence, namely the formation of zero-energy Majorana states. We then focus on topological Josephson junctions, which have been predicted to exhibit the {\it fractional Josephson effect}, first identified by Fu \& Kane \cite{Fu2008,Fu2009} as a signature of topological superconductivity.
\end{svgraybox}

\subsection{$p$-wave superconductivity in 2D and 3D topological insulators}
\label{subsec:1-1}

\runinhead{Proximity effect}

At the interface between a superconductor (S) of gap $\Delta$ and a normal (i.e. non superconducting) material (denoted N), the conversion of normal current into supercurrent (carried by Cooper pairs) and vice-versa is mediated by Andreev reflections. When an electron incident from the N side with energy $\epsilon<\Delta$ reaches the interface, a Cooper pair can be injected into the superconductor without breaking charge or energy conservation when combined with the retro-reflection of a hole with energy $-\epsilon$. This mechanism is called Andreev reflection, and is a key notion that governs the physics of two electronic states: 'superconducting' and 'normal' interacting by exchange of electrons at the interface. This quantum process is, at the nanoscale of a Josephson junction, not localized at the interface. Its extension is given by the so-called coherence length $\xi=\frac{\hbar v_F}{\Delta}$ ($v_F$ the Fermi velocity in the normal region) which measures, for a system without elastic scattering, how far correlations between paired electrons penetrate into the normal side. As a consequence, this length also naturally yields a {\it proximity effect} \cite{Klapwijk2004}, i.e. the typical distance over which superconductivity can be induced in a normal conductor by a superconductor located nearby.

\runinhead{Induced $p$-wave superconductivity}

When a nearby conventional superconductor induces superconductivity in a topological phase, the symmetries and properties of the induced superconductivity are deeply influenced by the peculiar transport properties in this phase. In a vast majority of experimentally relevant cases, superconductivity is induced by a conventional superconductor (Al, Pb, Nb, NbTiN), in which superconductivity arises from $s$-wave-paired electrons of opposite spins. 
In contrast, spin rotation symmetry is broken in 2D and 3D topological insulators, since electrons have to abide the so-called spin-momentum locking: electrons with opposite directions have opposite spins (in fact total angular momentum). Thus, topological phases give rise to induced "spinless" superconducting systems, since only one fermionic species (rather than two) is present and forms Cooper pairs. In other words, with spin rotation symmetry being broken in the topological phases, it must also be in the induced superconducting states and a so-called $p$-wave superconducting state with odd parity emerges \cite{Alicea2012,Beenakker2013,Aguado2017}.

\runinhead{Majorana bound states}

Such a $p$-wave superconductivity has several consequences, one of them being the existence of zero energy modes known as Majorana bound states. In $s$-wave superconductors, the Bogoliubov quasiparticle operators read $\hat\gamma_s= u\hat c_\uparrow^\dagger+v\hat c_\downarrow$ (where $\hat c_\sigma$ are electron annihilation operators of spin $\sigma=\uparrow,\downarrow$), for which $\hat\gamma_s\neq\hat\gamma_s^\dagger$. In contrast, thanks to the lifted spin degeneracy, a $p$-wave superconducting state allows for excitations such that $\hat\gamma_p= u\hat c^\dagger+v\hat c$. The famous condition for Majorana excitations $\hat\gamma_p=\hat\gamma_p^\dagger$ can thus be fulfilled for $u=v$. While forbidden in conventional $s$-wave superconductors, such states do exist in $p$-wave superconductors, and constitute the realization of Majorana fermions emerging in a condensed matter system. They naturally lie at zero-energy due to electron-hole symmetry, and localize at system boundaries and topological defects (such as vortices). One can show \cite{Alicea2012} that they support non-abelian statistics, and as such hold promise for exotic fundamental physics, and application to topologically protected quantum computation.

Given the properties of Majorana states, it is natural to investigate interfaces between a topological phase in a normal state and a superconductor, for example by tunnel spectroscopy. This route has led to intriguing observations of zero-bias anomalies in nanowires with strong spin-orbit coupling, in which similar physics should arise when a topological phase transition occurs under applied magnetic field along the axis of the nanowire \cite{Mourik2012,Albrecht2016}. In the remainder of the article, we focus on a different approach to topological superconductivity, namely the study of Josephson junctions in topological insulators (here HgTe, usually used for infrared detection technology). We in particular address how precursors of Majorana states alter the Josephson effect, and signal topological superconductivity.

\begin{figure}[t]
\sidecaption[t]
\includegraphics[width=11.5cm]{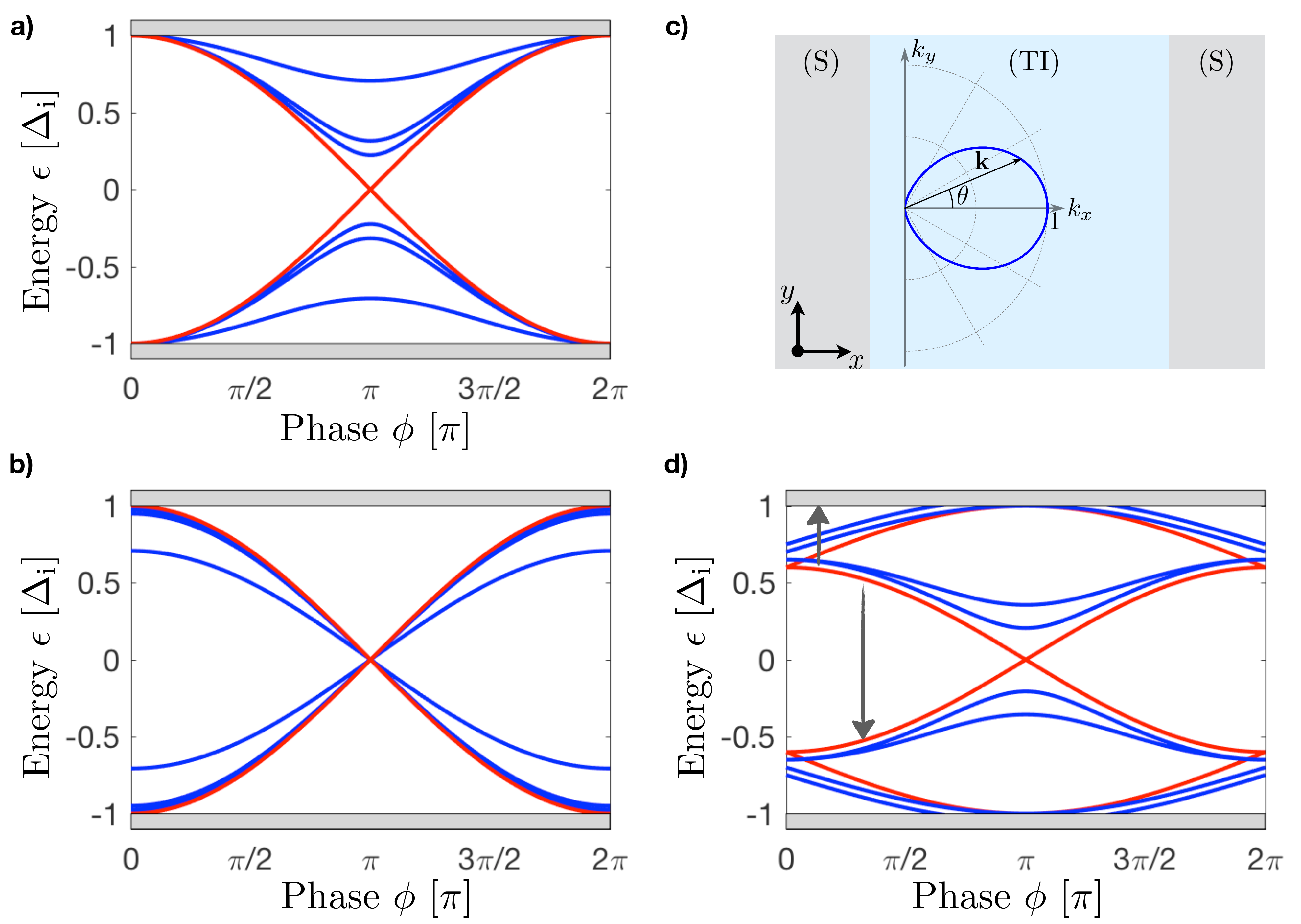}
\caption{Andreev bound states in 2D and 3D topological insulators --  a) Conventional Andreev bound states for different transmission coefficients $D$ : $D=1$ in red and $D=0.95,0.9,0.5$ in blue. b) Gapless Andreev bound states in $p$-wave superconductors for the same transmissions c) Polar plot of the transmission $D_\theta$ as a function of angle $\theta$ and scheme of a Josephson junction d) More realistic picture of an Andreev spectrum with one topological mode (red) and two non-topological modes (blue). Two possible relaxation mechanisms (ionization to the continuum, energy relaxation) are depicted as grey arrows.}
\label{fig:ABS}
\end{figure}

\subsection{Gapless Andreev bound states in 2D and 3D topological insulators}
\label{subsec:1-2}
The Josephson effect generically manifests itself as the occurrence of a phase-difference dependent non-dissipative supercurrent in a weak link between two superconductors. The nature of the weak link influences the properties of the supercurrent, which can thus serve as a probe of superconductivity. In mesoscopic systems, in general short compared to the phase correlation length, the supercurrent and its properties can be obtained by solving a scattering problem \cite{Beenakker1992}, with a weak link represented as a scattering matrix, and at each ends, boundary conditions set by Andreev reflections (together with normal reflection at the interface, because dissimilar materials have different electronic properties \cite{Blonder1982}). A number of resonant states thus form, called Andreev bound states, and their energies $\epsilon_n(\phi)$ depend on the superconducting phase difference $\phi$ between the two conductors. Major differences occur between the case of $s$- and $p$-wave superconductors \cite{Kwon2003}. We limit this discussion to the short junction limit, for which the length $L$ of the junction is much shorter that the coherence length $\xi$.

\runinhead{Conventional Andreev bound states}
In a conventional Josephson system between two $s$-wave superconducting reservoirs, Andreev bound states can generically be written as 
\begin{equation}
\epsilon_s(\phi)=\pm\Delta_{\rm i}\sqrt{1-D\sin^2\frac{\phi}{2}}
\end{equation}
where $D$ is the transmission of the weak link in the normal state, and $\Delta_{\rm i}$ the proximity induced gap. In Fig.\ref{fig:ABS}a,  $\epsilon_s(\phi)$ is represented, and shows that it is a $2\pi$-periodic function of $\phi$. An energy gap $2\Delta_{\rm i}\sqrt{1-D}$ is opened at $\phi=\pi$ for any $D\neq 1$. As will be seen below, the limit $D\to 1$ is singular: the spectrum becomes $\epsilon_s(\phi)=\pm\Delta_{\rm i}\left|\sin\frac{\phi}{2}\right|$, and the Andreev doublet is gapless, with $\epsilon_s(\pi)=0$. This regime is approached in superconducting atomic point contacts, that exhibit high transparencies $D>0.99$ \cite{Chauvin2006, Bretheau2013} and do not suffer from the use of dissimilar materials. Besides, in conventional systems, each Andreev doublet is in fact doubly degenerate as both spin species are active.

\runinhead{Gapless Andreev bound states in 2D topological insulators}
The Andreev bound states which form in topological weak links exhibit some remarkable differences as compared to the previous case, as shown in Fig.\ref{fig:ABS}b. Generically, in a 1D geometry, Andreev bound states forming between $p$-wave superconductors read \cite{Kwon2003,Kwon2004} :
\begin{equation}
\epsilon_p(\phi)=\pm\Delta_{\rm i}\sqrt{D}\sin\frac{\phi}{2}
\end{equation}
In the $s$-wave case, the transmission $D$ determines the avoided crossing at $\phi=\pi$. In contrast, in the $p$-wave case, an imperfect transmission $D<1$ opens a gap at $\phi=0$ between the continuum and Andreev states, with an amplitude $\Delta_{\rm i}(1-\sqrt{D})$.

Such bound states indeed describe the solutions of a weak link fabricated from a 2D topological insulator \cite{Fu2009}. There, spin polarized 1D edge channels are responsible for the electrical transport. In such a system, back-scattering is forbidden, as long as time-reversal symmetry is preserved. Then, one finds $D=1$, and the previous equation results in a unique $4\pi$-periodic Andreev doublet with:
\begin{equation}\epsilon_{2D}(\phi)=\pm\Delta_{\rm i}\sin\frac{\phi}{2}
\end{equation}
The degeneracy at $\phi=0$ is then a manifestation of time-reversal symmetry. Conversely, when time-reversal symmetry is broken, the transmission is reduced ($D<1$). The broken spin rotation symmetry results here in a lifting of the spin degeneracy: gapless Andreev bound states are not spin degenerate as opposed to their conventional counterparts. The two states of the doublet correspond to opposite fermion parities. The level crossing at $\phi=\pi$ is a manifestation of topology and is as such protected, and the gapless Andreev doublet (sometimes also called Majorana-Andreev bound states) can in fact be seen as the hybridization of two Majorana end states (see Section \ref{subsec:1-1}) bound at the two S-TI interfaces.

\runinhead{Superconducting Klein tunneling in 3D topological insulators} The Andreev energy spectrum in the 2D geometry of surface states in a 3D topological insulator is slightly richer. For a bar of width $W$, normal transport occurs through $N=W/\lambda_F$ modes, which results in $N$ Andreev doublets in a Josephson junction (see Fig.\ref{fig:ABS}c). For a wide junction, these doublets are indexed by the transverse momentum $k_y$ or, equivalently, by the angle $\theta$ such that $\cos\theta=\sqrt{1-\frac{k_y^2}{k_F^2}}$, and typically read as \cite{Tkachov2013} :
\begin{equation}
\epsilon_{3D}(\phi)=\pm\Delta_{\rm i}\sqrt{1-D_\theta\sin^2\frac{\phi}{2}}
\end{equation}
where $D_\theta$ is a $\theta$-dependent transmission. This generalized 2D transmissivity reflects the topological and Dirac nature of the charge carriers \cite{Tkachov2013,Tkachov2013a}, and can be written as :
 \begin{equation}
D_\theta=\frac{\cos^2\theta}{1-\frac{\sin^2\theta}{1+Z^2}}
\end{equation}
where $Z$ is a parameter characterizing the scattering (described here as a potential barrier). This system intrinsically hosts a single $4\pi$-periodic mode together with many $2\pi$-periodic ones which may simultaneously manifest in the Josephson response of the device. Indeed, a single topological Andreev doublet occurs at transverse momentum $k_y=0, \theta=0$ and is immune to back-scattering (thus has perfect transmission) as $D_{\theta=0}=1$ regardless of $Z$. On the contrary, a large number $\approx N$ of non-topological oblique modes ($k_y\neq0, \theta\neq0$) have lower transmissions $D_{\theta\neq0}<1$. In that sense, the topological protection of the zero mode constitutes a superconducting analogue to Klein tunneling. Though the Andreev bound states are not protected from scattering for $\theta=0$, they still feature the spin-momentum locking and may be called helical just as the topological surface states.

\runinhead{Beyond the short junction limit}

The preceding results are all obtained in the limit $L\ll\xi$, which for a single transport channel results in a unique Andreev doublet. Outside this regime, the situation is more complex, as more levels play a role in transport, with a typical level spacing of $\Delta \xi/L$. A schematic picture of a possible spectrum is presented in Fig.\ref{fig:ABS}d. In experiments, the exact spectrum is not known, and depends on parameters such as the length of the junction and the Fermi energy. Nevertheless, most features remain valid. In particular, there is in both 2D and 3D TIs a unique Andreev doublet with a protected level crossing at $\epsilon=0$ and $\phi=\pi$. It consequently exhibits $4\pi$-periodicity. We refer the reader to references \cite{Fu2009,Olund2012} for a more complete discussion.

\subsection{Fractional Josephson effect}
\label{sec:FracJE}

\subsubsection{Conventional and fractional Josephson effect}

\runinhead{Conventional Josephson effect}

An Andreev bound state of energy $\epsilon(\phi)$ carries a supercurrent, the amplitude of which is proportional to $\frac{\partial \epsilon}{\partial \phi}$. The so-called current-phase relation expresses the relation between the supercurrent $I_s$ and the superconducting phase difference $\phi$ between the two (undisturbed) macroscopic quantum phases of the superconductors on each side. It may be complicated when multiple ABS contribute, but its simplest expansion is :
 \begin{equation}
I_s(\phi)=I_c\sin\phi+\, {\rm higher\ harmonics}
\end{equation}
with $I_c$ the critical current of the junction, assumed to be a constant. The higher harmonics can in some cases represent an important contribution (for example high transmissions $D\to 1$). The main point here is however that it remains $2\pi$-periodic in $\phi$. When combined with the second Josephson equation $\frac{d\phi}{dt}=\frac{2eV}{\hbar}$, it is clear that a constant voltage $V$ gives rise to an oscillating current $I_s(t)=I_c\sin(2\pi f_{\rm J} t)$, with the conventional Josephson frequency $f_{\rm J}=\frac{2eV}{h}$, which is currently the basis for the voltage standard.

\runinhead{Fractional Josephson effect}
The presence of topologically protected Andreev bound states with $4\pi$-periodicity is expected to manifest itself as a {\it fractional Josephson effect} \cite{Kwon2003}, by modifying the equations describing the junctions. Indeed, the current-phase relation now fundamentally reads: 
 \begin{equation}
I_{2D/3D}(\phi)=I_{4\pi}\sin\frac{\phi}{2}+I_{2\pi}\sin\phi+\, {\rm higher\ harmonics}
\end{equation}
where $I_{4\pi}$ and $I_{2\pi}$ are two constants encoding the amplitude of the $4\pi$- and $2\pi$-periodic supercurrents. The Josephson supercurrent then oscillates with frequency $f_{\rm J}/2=\frac{eV}{h}$, hence the name {\it fractional Josephson effect} \cite{Kwon2003}.

The fractional Josephson effect should have two clear signatures. First, under constant DC bias, the oscillating Josephson current should result in an observable dipolar Josephson emission at $f_{\rm J}/2$, typically in the GHz range as $\frac{e}{h}\simeq \SI{0.5}{\giga\hertz\per\micro\volt}$ which can be measured and analyzed using rf techniques (see Section \ref{sec:emission}). Secondly, when phase locking occurs between the internal junction dynamics and an external microwave excitation at frequency $f$, 
Shapiro steps \cite{Shapiro1963} appear at discrete voltages given by $V_n = nhf/2e$, where $n$ is an integer step index. In the presence of a sizable $4\pi$-periodic supercurrent, an unconventional sequence of even steps (with missing odd steps), is expected, reflecting the doubled periodicity of the Andreev bound states \cite{Fu2009,Houzet2013,Badiane2013} (see Section \ref{sec:Shapiro}).

\subsubsection{Obstacles to the observation of the fractional Josephson effect}
\label{subsec:obstacles}
The above description must be carefully balanced out, as various phenomena can alter this simple picture.

\runinhead{Relaxation and thermodynamic limit} The previous signatures of the fractional Josephson effect are based on the hypothesis that the occupation number of the gapless Andreev levels remain constant, so that $I_{4\pi}$ is unchanged over the full duration of the experiment. Due to quasi-particle poisoning or ionization to the continuum (depicted as grey arrows in Fig.\ref{fig:ABS}d), the occupation of the $4\pi$-periodic fluctuates, which in turn affects the periodicity of the Josephson effect.

For a time-independent phase $\phi$, the occupation numbers reach the thermodynamical limit, and only the lower branches of Andreev bound states at $\epsilon\leq0$ are populated. The current is then $2\pi$-periodic. It can indeed be shown that at equilibrium $I(\phi)=e\frac{\partial \epsilon}{\partial \phi}\big(1-2f(\epsilon)\big)\propto\sin\frac{\phi}{2}\tanh(\frac{\Delta_i}{2kT}\cos\frac{\phi}{2})$, where $f$ is the Fermi-Dirac distribution function. This is a $2\pi$-periodic function, and is in fact identical to the expressions obtained for ballistic conventional Josephson junctions ($D\to 1$), and does not highlight the topological character of the induced superconductivity \cite{Kulik1977,Beenakker1991}.

Experiments relying on out-of-equilibrium dynamics are thus useful to provide evidence for the existence of gapless $4\pi$-periodic Andreev bound states on time scales shorter than the equilibration time. Only on such short time scales can one in principle observe doubled Shapiro steps, or the anomalous Josephson emission at half the Josephson frequency $f_{\rm J}/2$. We focus in the next sections on experiments focusing on dynamics the GHz range. We also refer the reader to several works on the effects of relaxation mechanisms on the signatures of topological superconductivity \cite{Fu2009,Badiane2011,SanJose2012,Pikulin2012,Lee2014}.

\runinhead{Coupling to the continuum}
Second, it is important to notice that time-reversal symmetry should in principle impose a Kramers degeneracy point at $\varphi=0,2\pi,...$. There, the gapless Andreev bound states are either connected to other Andreev states or to the bulk continuum \cite{Fu2009,Beenakker2013a}. Bulk quasiparticles are then produced as $\phi$ is adiabatically advanced. This naturally leads to enhanced relaxation at these points, and suppresses the dissipationless and $4\pi$-periodic character of the supercurrent, thus restoring a $2\pi$-periodicity. 

Surprisingly, these degeneracies are modified when electron-electron interactions are taken into account in a many-body picture. Then, the many-body Andreev spectrum is reorganized and gives rise to an effective $8\pi$-periodic supercurrent when combined with electron interaction \cite{Zhang2014,Peng2016,Hui2016,Vinkler2017,Pedder2017}. In that case, instead of the $4\pi$-periodic fractional Josephson effect, one may expect to observe an $8\pi$-periodic Josephson effect, with Shapiro steps only visible with index $n\equiv 0\bmod 4$, or emission at a quarter of the Josephson frequency $f_{\rm J}/4$. 

\runinhead{Landau-Zener transitions}Another important {\it caveat} is the possibility of Landau-Zener transitions (LZT) between Andreev bound states near an avoided level crossing. When the voltage $V$ or equivalently the frequency $f$ are sufficiently high, LZT can mimic an effective $4\pi$-periodic Josephson effect, while the spectrum of Andreev states remains gapped, with only a small avoided crossing at $\phi=\pi$ \cite{Dominguez2012,Pikulin2012,Sau2012}. Such LZTs have previously been observed in single Cooper pair transistors \cite{Billangeon2007} and can in principle be distinguished from an intrinsic fractional Josephson effect by a strong voltage dependence of the emission or Shapiro step features.\\
\newline
In Section \ref{sec:TopoOrigin}, special attention will be given to assessing the topological origin of the observed fractional Josephson effect and the role of the aforementioned mechanisms.

\section{HgTe-based Josephson junctions and experimental techniques}

\begin{svgraybox}
Here, we first briefly introduce the reader to the geometry, fabrication and basic properties of the devices. In particular, we show how we estimate parameters such as the induced gap $\Delta_{\rm i}$ and the coherence length $\xi$. Then, we describe the setups operated to measure the response to ac excitations (Shapiro steps) or capture Josephson emission.
Given their high mobilities and low intrinsic electron densities, we argue that HgTe-based 2D and 3D topological insulators thus appear as ideal base material to fabricate topological Josephson junctions and observe the manifestations of topological superconductivity. Finally we conclude this section with a presentation of the experimental setups which allow for a simple, fast and reliable measurement of the devices.
\end{svgraybox}

\subsection{Fabrication of HgTe-based Josephson junctions}

The junctions are fabricated from epitaxially grown layers of HgTe on a CdTe substrate for which the mobility and carrier density are evaluated from a Hall-bar produced separately prior to the fabrication of Josephson junctions. Both 2D and  3D topological insulators can serve as weak links in Josephson junctions. However, these early devices made of 3D topological insulators suffer from lower mobilities caused by the absence of protective capping layer (CdHgTe), and from the absence of gate electrode to tune the electron density \cite{Maier2012,Oostinga2013,Maier2015}. We briefly review below the main characteristics of the different devices.

\begin{figure}[t]
\sidecaption[t]
\includegraphics[width=11 cm]{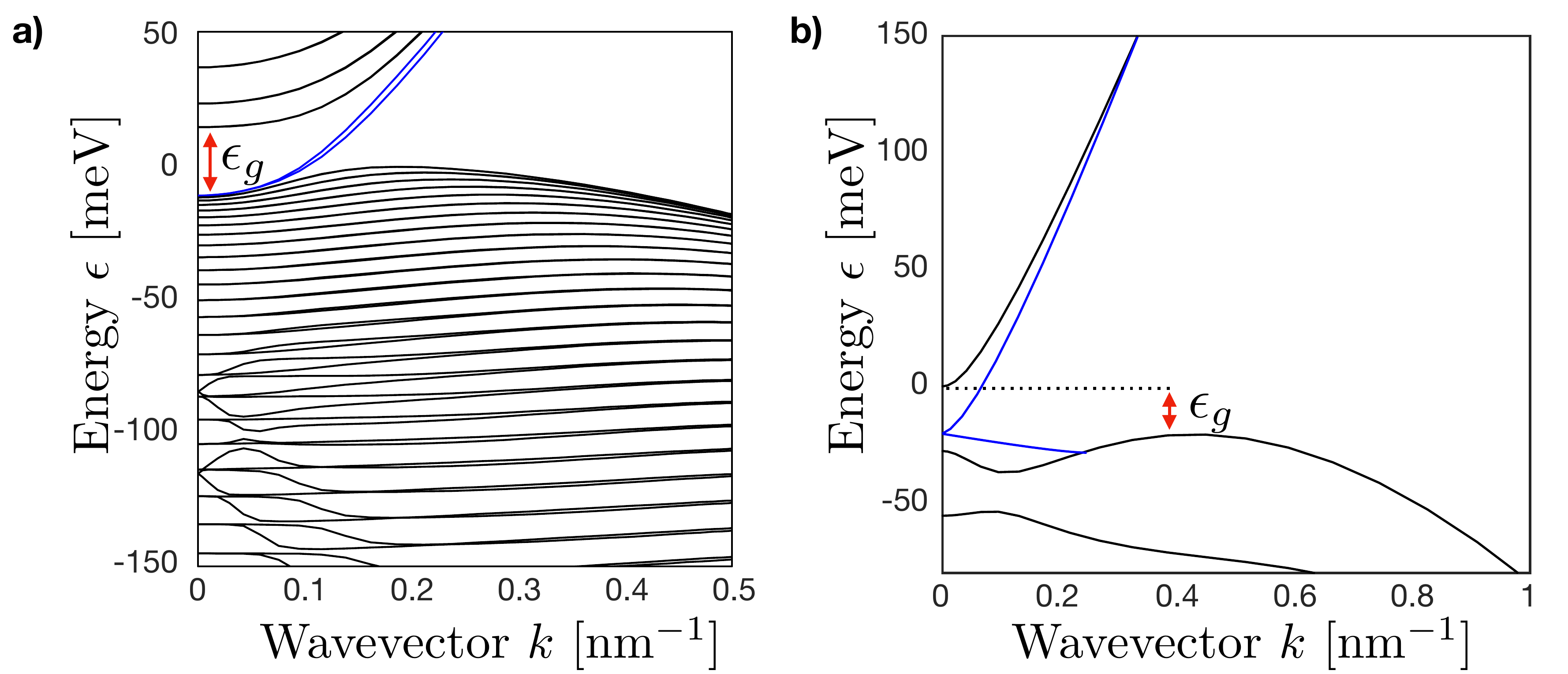}
\caption{Band structures of HgTe-based 2D and 3D topological insulators from $k.p$ simulations -- a) Strained HgTe layer of \SI{70}{\nano\meter} : bulk 3D states are pictures in black, while 2D topological surfaces states appear in blue b) HgTe quantum well of \SI{7.5}{\nano\meter}: bulk 2D states appear in black, while 1D topological edge states (not simulated) are indicated in blue.}
\label{fig:BandStructures}
\end{figure}

\runinhead{Strained HgTe as a 3D topological insulator} The 3D topological insulators (TI) are obtained from coherently strained undoped HgTe layers of \num{60} to \SI{90}{\nano\meter} thickness. The band inversion of HgTe enforces the existence of topological surface states, while strain opens a gap ($\epsilon_g\simeq \SI{20}{\milli\electronvolt}$) in the bulk of the material \cite{Fu2007a}. A typical band structure is shown in Fig.\ref{fig:BandStructures}a for a strained \SI{70}{\nano\meter}-thick HgTe layer. In previous works, we have proven the high quality of the topological states in this material \cite{Bruene2011, Bruene2014} and notably they entirely dominate electron transport up to very large electron densities \cite{Inhofer2017,Tchoumakov2017}. In the many devices tested, we find typical densities of $n_{3D}= 3 - \SI{7e11}{\per\centi\meter\squared}$ and mobilities of $\mu_{3D}=1 -\SI{3e4}{\centi\meter\squared\per\volt\per\second}$. From these values, we calculate a mean free path of $l_{3D}\simeq \SI{200}{\nano\meter}$.

\runinhead{HgTe quantum wells as a 2D topological insulator} New lithography processes have enabled to fabricate Josephson junctions from thin quantum wells of HgTe, sandwiched between barrier layers of Hg$_{0.3}$Cd$_{0.7}$Te grown on a CdZnTe substrate \cite{Bernevig2006}. As depicted on Fig.\ref{fig:BandStructures}b, for thicknesses above a critical thickness $d_c\simeq\SI{6.3}{\nano\meter}$, topological edge channels are expected between the conduction and valence band, separated by a small gap $\epsilon_g$, with $\epsilon_g\simeq 10 - \SI{30}{\milli\electronvolt}$ depending on growth parameters. The existence of topological edge states has been proven via transport measurements \cite{Konig2007, Roth2009, Bruene2012} and scanning-SQUID imaging \cite{Nowack2013}. 
The typical densities are $n_{2D}= 1 - \SI{5e11}{\per\centi\meter\squared}$ with mobilities $\mu_{2D}=3 -\SI{5e5}{\centi\meter\squared\per\volt\per\second}$. As a consequence, 2D devices are expected to have a larger mean free path of $l_{2D}\simeq \SI{2}{\micro\meter}$ compared with the thick 3D layers.

Remarkably, it is possible to grow thinner quantum wells ($d<d_c$) that do not exhibit a band inversion, and consequently do not host any topological edge channels. Outside of the gap region, such layers are extremely similar to thick quantum wells, and exhibit the same typical densities and mobilities. They are as such ideal reference samples to benchmark the experimental techniques and observations in a trivial system. We will refer to such a reference sample in the rest of the article.

\begin{figure}[t]
\sidecaption[t]
\includegraphics[width=11 cm]{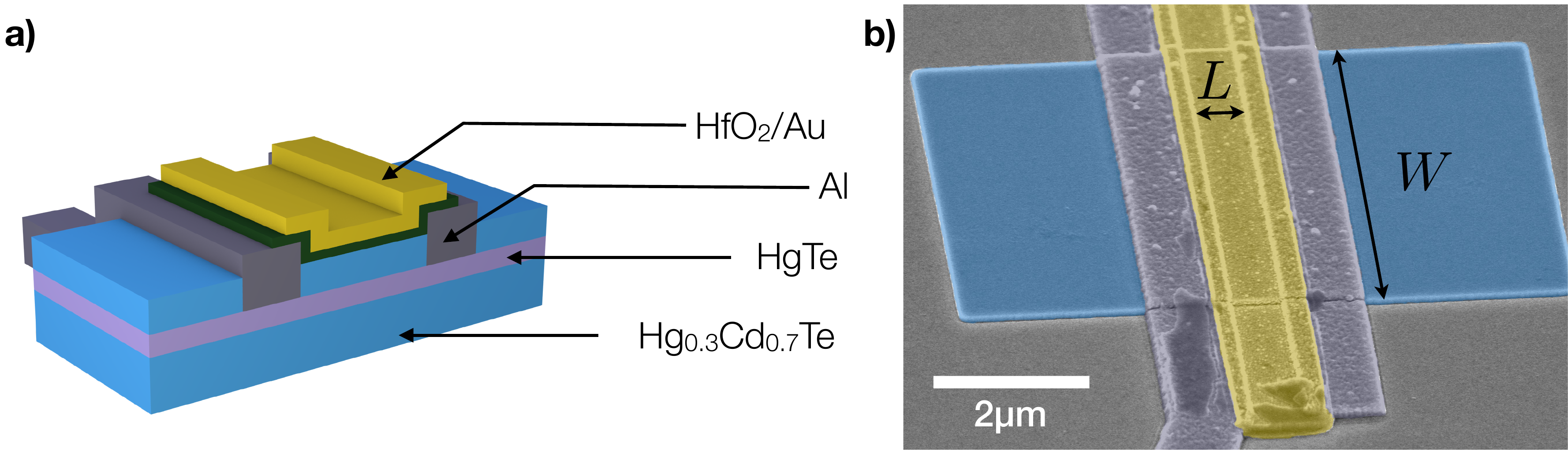}
\caption{Experimental realization of a HgTe-based topological Josephson junction -- Artist view (a) and colorized SEM picture (b) of a junction. The HgTe 2D topological insulator (in mauve) is sandwiched between two layers of Hg$_{0.3}$Cd$_{0.7}$Te (in blue). The Al superconducting contacts are in dark purple while the gate is in yellow and rests on a thin dielectric layer of HfO$_2$ (dark green). Devices realized on 3D TIs are similar, without the top-gate and the Hg$_{0.3}$Cd$_{0.7}$Te protective top layer.}
\label{fig:Devices}
\end{figure}

\runinhead{Geometry of the Josephson junctions} The layout of the Josephson junctions is shown in Fig.\ref{fig:Devices} and is similar for both 2D TIs, 3D TIs and reference samples (apart from the absence of the gate and protective cap layer on the 3D sample). A rectangular mesa of HgTe is first defined. First the oxide and capping layers are etched, before superconducting contacts are deposited on the HgTe layer. Niobium has been sputtered on 3D TI samples, while Al is used with standard evaporation techniques on 2D TI and reference samples. Upon the latter, a metallic gate electrode of Au is added between the Al contacts on an HfO$_2$ dielectric layer grown via Atomic Layer Deposition (ALD) to control the electron density. 

The superconducting stripes have a width of \SI{1}{\micro\meter}. The HgTe mesa has a width $W=2-\SI{4}{\micro\meter}$, determining the width of the weak link. In 3D topological insulators, it is advantageous to narrow down the mesa so as to reduce the number of bulk modes with $k_y\neq0$, while in contrast a large mesa reduces the overlap of edge channels on opposite edges in 2D samples \cite{Zhou2008}. The length of the junctions have been varied between $L=\SI{200}{\nano\meter}$ and \SI{600}{\nano\meter}. 

\subsection{Basic properties of HgTe-based Josephson junctions}

Before moving to measurements specific to the fractional Josephson effect, the study of the $I$--$V$ curves of the junctions under DC bias provides some information on the microscopic parameters of the junctions, that we review in this section. 

\begin{figure}[t]
\sidecaption[t]
\includegraphics[width=11.8 cm]{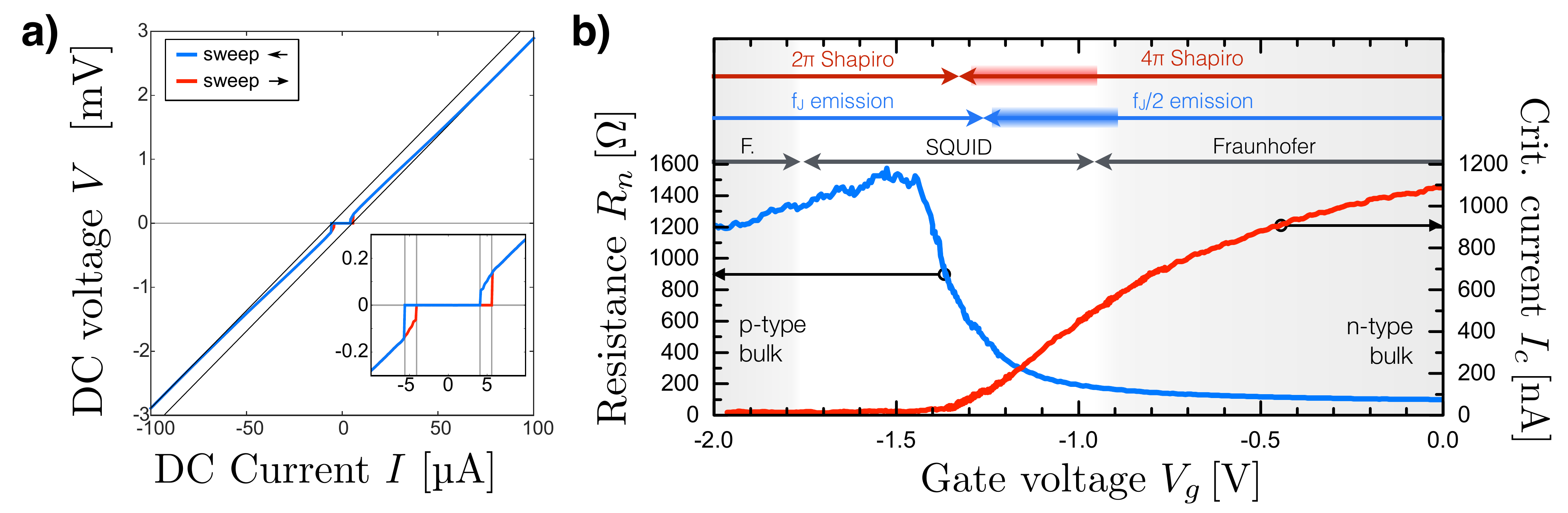}
\caption{DC characterization of HgTe-based Josephson junctions -- a) Junction with a 3D topological insulator : $I$--$V$ curve of a Josephson junction fabricated from a 3D TI, taken at base temperature $T\simeq \SI{30}{\milli\kelvin}$. The asymptotes (grey solid lines) do not cross the origin, emphasizing the presence of an excess current. (Inset) Detailed view of the $I$--$V$ curve, that exhibits hysteresis between the upward and downward sweep direction. b) Junction with a 2D topological insulator : critical current $I_c$ (red line) and normal state resistance $R_n$ (blue line) as a function of gate voltage $V_g$. The red, gray and blue arrows summarize the ranges where we observe the anomalous Josephson effect properties. The arrows are thicker where the emission at $f_{\rm J}/2$ and the even sequences of Shapiro steps are the most visible. Panel a adapted from \cite{Wiedenmann2017}}
\label{fig:JJCharac}
\end{figure}

\runinhead{$I$--$V$ curve of Josephson junctions} As mentioned earlier, the junctions based on 3D TIs do not have a gate, and their electron density is such that the number of transport modes lies typically between $N=50$ and 200, depending on the sample width (with variations of about 30\% for a given dimension). A typical $I$--$V$ curve is presented in Fig.\ref{fig:JJCharac}a, and shows the expected behavior of a Josephson junction with a critical current of $I_c\simeq\SI{5}{\micro\ampere}$. It exhibits hysteresis, as commonly reported \cite{Oostinga2013}. We believe that the hysteresis is an intrinsic ingredient of mesoscopic Josephson devices, which reflects the difference in the Josephson current amplitude in the static (DC) case compared to the dynamic (AC) case \cite{Galaktionov2010,Antonenko2015}.

In all devices, an excess current in the $I$--$V$ curve is clearly visible, with an asymptote which does not go through the origin but is shifted towards higher currents. This excess current reflects the high probability of Andreev reflections in an energy window near the superconducting gap \cite{Blonder1982,Klapwijk1982}, and underlines the high quality and reproducibility of our devices in line with previous observations \cite{Oostinga2013,Maier2015,Sochnikov2014}.

An important parameter of the junction is the amplitude of the induced superconducting gap $\Delta_{\rm i}$. We have resorted to the study of the temperature dependence of the critical current and obtained estimates on the order of $\Delta_{\rm i}=100-\SI{350}{\micro\electronvolt}<\Delta_{\rm Nb}$, but with a large uncertainty given the lack of adequate theories \cite{Tkachov2013}. The relevant coherence length for the quasi-ballistic weak link is then estimated $\xi=\sqrt{\frac{\hbar v_{\rm F}l}{\pi\Delta_{\rm i}}}$ in the range of 250 to \SI{550}{\nano\meter}, and is compatible with our observations of the decay of $I_{\rm c}$ with length. The 3D junctions are consequently in an intermediate regime $l\sim\xi\sim L$, which is particularly hard to model, as the junctions reach neither the short (corresponding to $L\ll\xi$) nor the ballistic ($L\gg l$) limit.

Similarly, the study of $I$--$V$ curves for 2D weak links that $\Delta_{\rm i}\simeq\SI{40}{\micro\electronvolt}$, compatible with the gap of the Al contacts ($\Delta_{\rm Al}\simeq \SI{100}{\micro\electronvolt}$). Our junctions are consequently in an intermediate length regime $L\sim\xi$, given the estimated coherence length $\xi~\simeq~\SI{600}{\nano\meter}$, but reach the ballistic limit $L\ll l$ owing to the large mean free path $l>\SI{2}{\micro\meter}$.

\runinhead{Mapping to the band structure in 2D topological insulator junctions} 
The presence of a gate however enables to vary the electron density and to identify different transport regimes from the normal state resistance $R_n$ and the critical current $I_c$. In agreement with the band structure presented in Fig.\ref{fig:BandStructures}b, we distinguish three regimes. For gate voltages between $V_g=\SI{-1.1}{\volt}$ and \SI{0}{\volt}, $R_n$ is low (below \SI{300}{\ohm}) and $I_c$ is large (above \SI{200}{\nano\ampere}). This signals the $n$-conducting regime, with a high mobility and high electron density in the plane of the junction. For gate voltages below $V_g=\SI{-1.7}{\volt}$, the normal state resistance tends to decrease slowly, indicating the $p$-conducting regime with a significantly lower mobility. The critical current $I_c$ lies however below \SI{50}{\nano\ampere}. Between these two regimes, a peak in $R_n$ (maximum around $\SI{1.5}{\kilo\ohm}$) and the quasi-suppression of $I_c$ indicates the region where the QSH edge states should be most visible. The peak value of $R_n$ is however much lower than the quantized value $h/2e^2\simeq\SI{12.9}{\kilo\ohm}$, and underlines the presence of residual bulk modes in the junction \cite{Hart2014}. 

We here like to point out that the trivial narrow quantum wells used for reference samples exhibit a similar gate dependence for $I_c$ and $R_n$. On the short length of the junctions, the gapped region (intermediate gate range) is not strongly insulating, maybe due to percolation transport due to disorder and residual charge puddles. In contrast, the gap region is observed to be strongly insulating on larger devices such as the Hall bars used for characterization of the layer properties. 

The observations on 2D topological insulator junctions can be further validated by scrutinizing the response to a magnetic field perpendicular to the plane of the junction. Then the superconducting phase difference $\phi$ becomes position dependent \cite{Tinkham2004}, a property which helps revealing the spatial supercurrent distribution through modulations of the critical current $I_c$. When the electron density is high and the current flows uniformly in the 2D plane of the quantum well, the junction exhibits a conventional Fraunhofer pattern of the critical current versus magnetic field, that rapidly decays as the magnetic field increases. This indicates the $n$- and $p$-conduction regimes, respectively for $V_g>\SI{-0.9}{\volt}$ and $V_g<\SI{-1.8}{\volt}$. In contrast the diffraction pattern is similar to that of a (DC) SQUID for $V_g$ between $\SI{-1.8}{\volt}$ and $V_g=\SI{-0.9}{\volt}$. It demonstrates that a large fraction of the supercurrent flows along the edges of the sample \cite{Hart2014}, as expected in the presence of QSH edge channels. We refer the reader to Ref.\cite{Bocquillon2016} for a detailed discussion.

\subsection{Experimental setups}

\begin{figure}[t]
\sidecaption[t]
\includegraphics[width=11 cm]{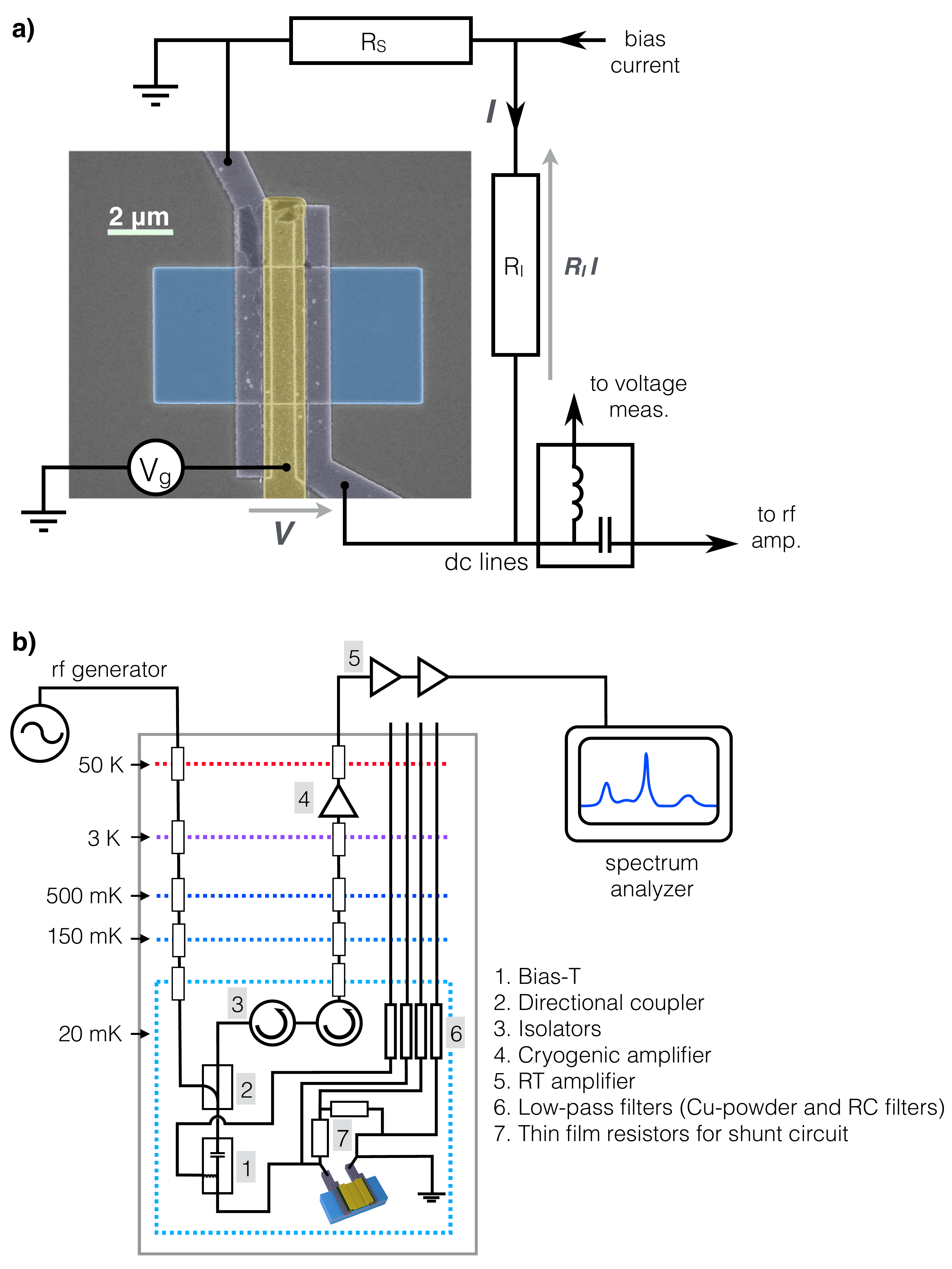}
\caption{Simplified setups for the Josephson emission measurements -- a) DC bias circuit, with a shunt resistance $R_S$ to achieve a stable voltage bias and a resistance $R_I$ to access the voltage $V$ and current $I$. The rf signal is coupled to the amplification scheme via a bias-tee b) Microwave amplification and detection setup, with cryogenic HEMT and room-temperature microwave amplifiers and a directional coupler to allow measurements of Shapiro steps and Josephson emission with a unique setup. Adapted from \cite{Deacon2017}.}
\label{fig:Setups}
\end{figure}

\runinhead{Microwave setup for Josephson emission} The simplest and most direct technique to measure Josephson radiation consists in measuring, after amplification, the emission spectra of the junctions with a spectrum analyzer. To this end, the junction is connected to a coaxial line and decoupled from the dc measurement line via a bias-T (see Fig.\ref{fig:Setups}b). The rf signal is then amplified at both cryogenic and room temperature before being measured with a spectrum analyzer. The commercial rf components used in the readout line for our measurements limit the frequency range of detection to approximately 2-\SI{10}{\giga\hertz}.
In spirit, this approach is similar to early measurements of Josephson emission using narrow-band resonant cavities \cite{Yanson1965,Pedersen1976}, but with extended bandwidth thanks to microwave cryogenic amplifiers \cite{Schoelkopf1995}. It contrasts with measurements using tunnel junctions as detectors \cite{Deblock2003,Onac2006,Billangeon2007}. This technique has the advantage of even wider bandwidths, but rely on the numerical deconvolution of modified $I$--$V$ characteristics of Al tunnel junctions, the interpretation of which being difficult in some cases \cite{Laroche2017}.

\runinhead{Microwave excitation for Shapiro steps} The formation of Shapiro steps can be easily observed in the dc $I$--$V$ characteristics when a Josephson junction is under rf excitation. The latter can be provided using either an open-ended coaxial cable (the end of which is placed a few millimeters from the sample), or by micro-bonding a lead to a microwave line, for example via a directional coupler (see Fig.\ref{fig:Setups}) to enable measurements of Josephson emission and Shapiro steps with a unique setup. In both geometries, frequencies in the range of 0 to 15 GHz are easily accessible, but the rf power supplied to the sample can not easily be calibrated.

\runinhead{DC bias circuit} An essential requirement for these measurements is to obtain a stable DC biasing of the junctions. We found that instabilities and hysteretic behavior occur at low voltages \cite{Wiedenmann2016,Bocquillon2016} and therefore employ a small resistive shunt $R_S$ (between 1 and \SI{50}{\ohm}) to enable a stable voltage bias. A small resistance $R_I$ in series with the junction can be included to enable the measurement of the current $I$ through the junction (Fig.\ref{fig:Setups}a). With an adequately filtered fridge, we have been able to observe stable emission features down to about \SI{1}{\giga\hertz} and Shapiro steps down to circa \SI{500}{\mega\hertz}.

\section{Experimental observation of the fractional Josephson effect}

\begin{svgraybox}
In this section, we review the first observation of the fractional Josephson effect in the topological Josephson junctions based on HgTe, both in 2D and 3D topological insulators. We juxtapose the results obtained in the two systems to highlight their similarities, and compare them to the reference situation provided by a quantum well in the trivial regime.
\end{svgraybox}

\subsection{Observation of Josephson emission at $f_{\rm J}/2$}
\label{sec:emission}
\subsubsection{Conventional Josephson emission}

\begin{figure}[t]
\sidecaption[t]
\includegraphics[width=11.1 cm]{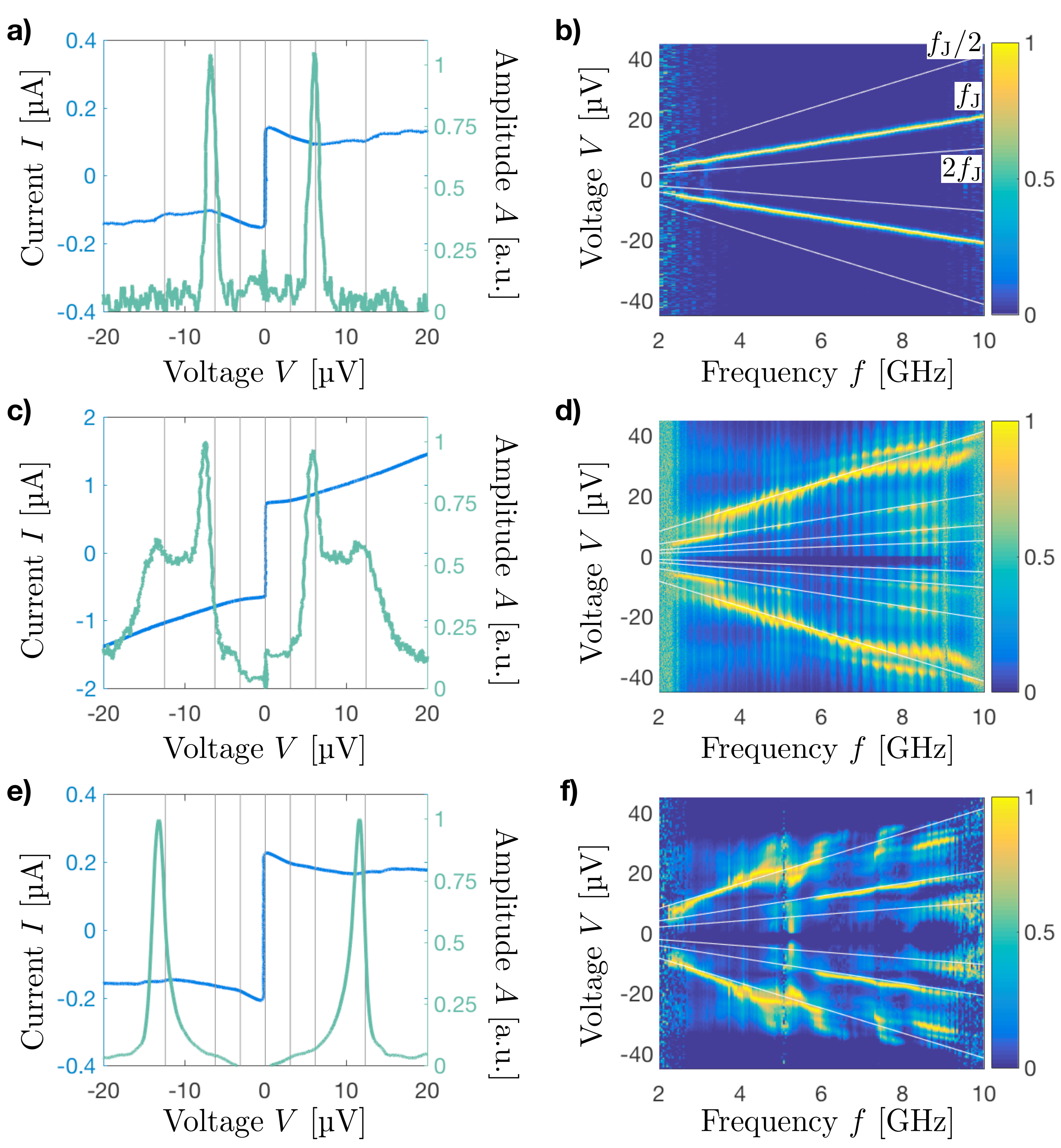}
\caption{Josephson emission in trivial and topological Josephson junctions -- In the first column (a,c,e), an emission spectrum (amplitude $A$) taken at $f=\SI{3}{\giga\hertz}$ is plotted in green, alongside with the $I$--$V$ curve of the device under consideration (depicted in blue). In the second column (b,d,f), the collected microwave amplitude $A$ is presented as a colorplot, as a function of frequency $f$ and voltage $V$. For better visibility, the data is normalized to its maximum for each frequency, and white guidelines indicate the $f_{\rm J},f _{\rm J}/2$ and $2f_{\rm J}$ lines (see panel b). The topologically trivial quantum well is shown in the first line (a, b) while the second (c, d) and third (e, f) show respectively the 3D topological weak links and the 2D ones (taken at $V_g=\SI{-0.55}{\volt}$). In panel c, the current $I$ is actually the total bias current, sum of the current in the shunt resistor $R_s$ and in the junction branches (Fig.\ref{fig:Setups}a). The second resistor $R_I$ has been here suppressed. This simplified circuit does not enable a proper measurement of the current in the junction only, but provides a correct readout of voltage $V$.}
\label{fig:DataEmission}
\end{figure}

We first focus on the investigation of Josephson emission. As a reference, we first discuss the case of the narrow HgTe quantum well in the trivial regime (Fig.\ref{fig:DataEmission}, first line). In the first panel (Fig.\ref{fig:DataEmission}a), the blue line indicates the $I$--$V$ curve of the device. At zero bias on the junction, a background noise originating from black body radiation and parasitic stray noise from the environment is observed. It is subtracted from all measurements to isolate the contribution of the junction. When the junction is biased, a finite voltage $V$ develops and the contribution of the junction appears, and is plotted as a green line. The observed peak in the emission at $V\simeq \SI{6}{\micro\volt}$ corresponds to the matching of the detection frequency $f_d$ with the Josephson frequency $f_{\rm J}\simeq f_d=\SI{3}{GHz}$. Sometimes, a second peak is observed at half this voltage, indicating a weak second harmonic at $2f_{\rm J}$. The proportionality of $f_{\rm J}$ with $V$ can be further verified by varying the detection frequency $f_d$, as shown in Fig.\ref{fig:DataEmission}b. A single emission line is observed, and fits perfectly with the theoretical prediction $f_{\rm J}=\frac{2eV}{h}$. This constitutes the expected signature of the conventional Josephson effect, as already observed in the early days of Josephson physics \cite{Yanson1965,Pedersen1976}. Besides, by varying the critical current $I_c$ (with the gate voltage $V_g$), we verify that the amplitude $A$ of the collected signal is proportional to $A\propto I_c$ with good agreement \cite{Likharev1986}, and consequently reaches its minimal amplitude in the gap region.

\subsubsection{Fractional Josephson emission}

In strong contrast, the junctions fabricated from 2D and 3D topological insulators reveal a strong emission peak at half the Josephson frequency $f_{\rm J}/2$  (Fig.\ref{fig:DataEmission}, second and third lines). This constitutes the most direct evidence of a $4\pi$-periodic supercurrent flowing in these junctions \cite{Badiane2011}.

\runinhead{Emission line at $f_{\rm J}/2$} The observation of emission at half the Josephson frequency is illustrated for $f=\SI{3}{\giga\hertz}$ in panels \ref{fig:DataEmission}c and \ref{fig:DataEmission}e. As seen for the 3D TI, the emission at $f_{\rm J}/2$  is sometimes concomitant with emission at $f_{\rm J}$, depending on frequency or gate voltage. We detail below these aspects.

Besides, a recurring observation is that the linewidth of the emission line at $f_{\rm J}/2$ is also larger (by up to a factor 10) than the conventional line at $f_{\rm J}$. For instance in the quantum wells, both the topological and trivial devices exhibit a line at $f_{\rm J}$ with a typical width of $\delta V_{2\pi}\simeq 0.5 - \SI{0.8}{\micro\volt}$. The linewidth at $f_{\rm J}/2$ exhibits values over a larger range, with widths in the range $\delta V_{4\pi}\simeq 0.5 - \SI{8}{\micro\volt}$. 

\runinhead{Dependence on frequency} We first discuss the data collected on the 3D topological insulator. In this device, we observe that the $f_{\rm J}/2$ line is dominant for a large range of voltages (12 to \SI{35}{\micro\volt}) or equivalently of frequencies (3 to \SI{9}{\giga\hertz}). Outside that range, the conventional emission at $f_{\rm J}$ dominates. The data collected on the 2D topological insulator (Fig.\ref{fig:DataEmission}b) is measured in the vicinity of the quantum spin Hall regime. In that case, the emission is clearly dominated by the $f_{\rm J}/2$ line below $f=\SI{5.5}{\giga\hertz}$, before the conventional line at $f_{\rm J}$ is recovered. We propose an interpretation of the influence of frequency in Section \ref{sec:TopoOrigin}.

In both cases, the emission lines deviate from the expected emission lines and a more complex structures with broadening, and multiple peaks is observed. We have identified resonant modes in the electromagnetic environment. In a dynamical Coulomb blockade situation, they possibly alter the emission spectrum in that range (see \cite{Deacon2017}) and are known to result in emission at $f_{\rm J}/2$. Nevertheless, they cannot solely explain the fractional Josephson effect. Indeed, these are second-order processes in $R_n/R_K$ (with $R_K = \frac{h}{e^2}$) and the amplitude of the $f_{\rm J}/2$ line always remains of lesser amplitude than standard emission at $f_{\rm J}$ \cite{Holst1994,Hofheinz2011}. 

\runinhead{Dependence on gate voltage (2D TI)}

As mentioned earlier, devices fabricated from 2D topological insulators enable to tune the electron density via the gate voltage $V_g$. We have clearly observed in these devices three regimes in the emitted power, that correlate with the expected band structure, as reported in Fig.\ref{fig:JJCharac}b. When the gate voltage is above $V_g>\SI{-0.4}{\volt}$, we observe that emission occurs for $f_{\rm J}/2$ is observed at low frequency, but the conventional line is recovered and dominates above typically \SI{5}{\giga\hertz}. These observations suggest transport in the conduction band of the quantum well, where gapless Andreev bound states have been seen to coexist with $n$-type conventional states, in agreement with previous observations and predictions \cite{Bocquillon2016,Dai2008}. However, in a narrower gate range $\SI{-0.8}{\volt}<V_g<\SI{-0.6}{\volt}$, one observes almost exclusively emission at half the Josephson frequency $f_{\rm J}/2$ up to very high frequencies (circa 8-\SI{9}{\giga\hertz}). We attribute this observation to the quantum spin Hall regime, where edge states are the dominant transport channel. For $V_g<\SI{-0.8}{\volt}$, the Josephson radiation at $f_{\rm J}/2$ is weakly visible, which suggests that the gapless Andreev modes more rapidly hybridize with bulk $p$-type conventional modes of the valence band. The overall gate voltage dependence is consistent with the expected band structure of a quantum spin Hall insulator, as presented in Fig.\ref{fig:BandStructures}b, but a quantitative description of the features remains difficult due to the observed irregularities in the emission lines.

\subsection{Observation of even sequences of Shapiro steps}
\label{sec:Shapiro}

\subsubsection{Conventional Shapiro response}

\begin{figure}[t]
\sidecaption[t]
\includegraphics[width=11.8 cm]{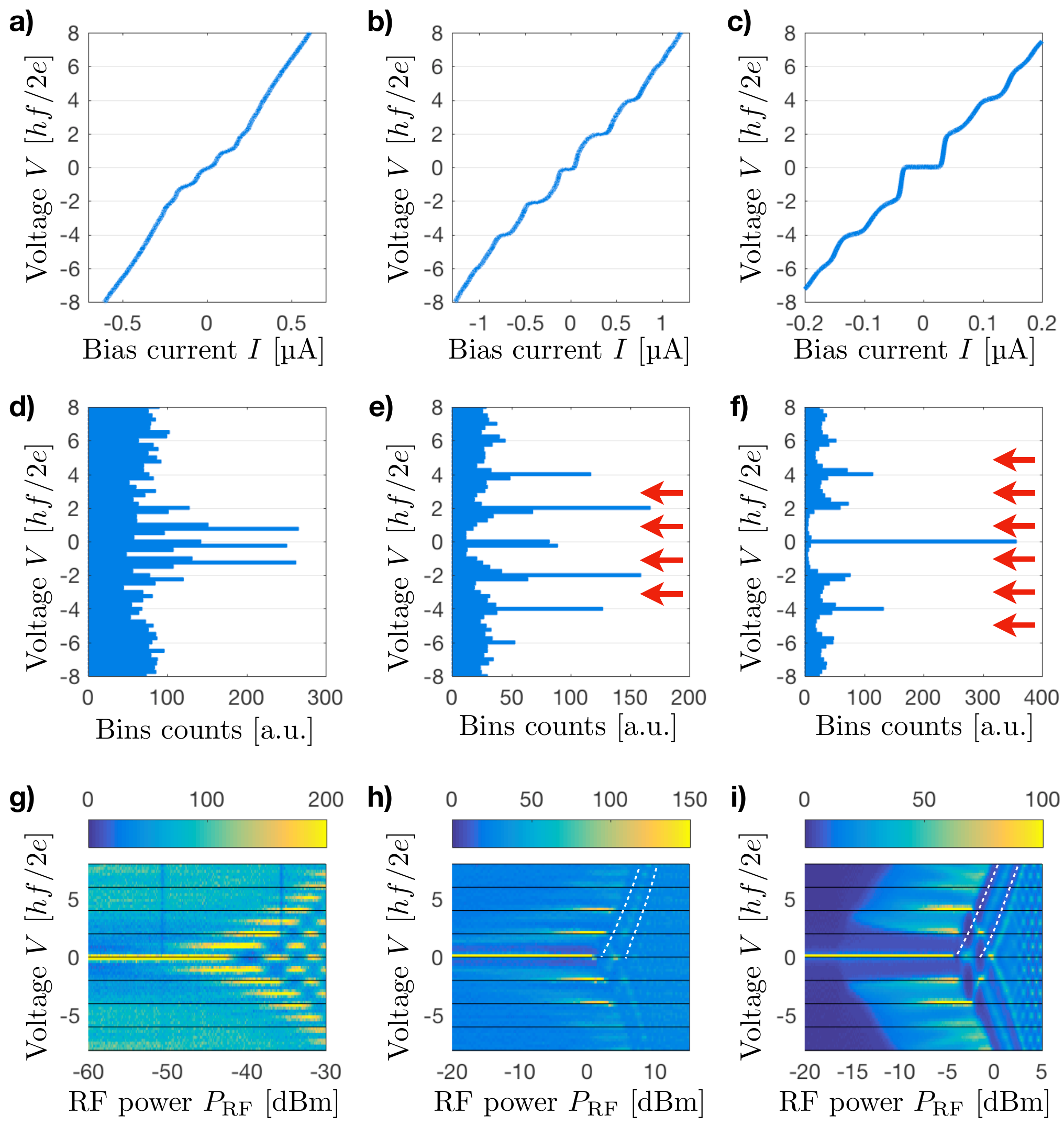}
\caption{Shapiro steps in trivial and topological Josephson junctions -- In the first line (a,b,c), typical $I$--$V$ curves the device under consideration are plotted. They exhibit Shapiro steps, the step index of which can be read from the normalized voltage $V$. In the second line (d,e,f), the histograms corresponding to the previous $I$--$V$ curves are shown as barplots and highlight the vanishing of odd steps (indicated by red arrows) in 2D and 3D topological weak links. Finally, the last line (g,h,i) presents colorplots obtained from the previous histograms, as a function of voltage $V$ and microwave power $P_{\rm RF}$.
The first column (a,d,g) shows data from a reference non-topological device (for $f=\SI{5.64}{\giga\hertz}$), the second column (b,e,h) from a 3D topological insulator ($f=\SI{1}{\giga\hertz}$), and the last column (c,f,i) from a 2D topological insulator ($f=\SI{1}{\giga\hertz}$). }
\label{fig:DataShapiro}
\end{figure}

We now turn to the second signature of the fractional Josephson effect, namely the observation of even sequences of Shapiro steps. Under microwave excitation at frequency $f$, the presence or absence of steps can be observed directly in the $I$--$V$ curves, but are conveniently highlighted by binning the measurement data according to the voltage. For $V_n=nhf/2e$ with $n$ integer, Shapiro steps then appear as peaks in the bin counts, their amplitude then reflects the current range over which the DC voltage stays fixed. In Fig.\ref{fig:DataShapiro}, we present $I$--$V$ curves for a given power of microwave irradiation (a,b,c), the histograms resulting from the binning as bar plots (d,e,f) and finally the same histograms in a colorplot function of voltage $V$ and microwave power $P_{\rm RF}$.

We first concentrate on the trivial Josephson junction as a reference situation (first column). Panels \ref{fig:DataShapiro}a and \ref{fig:DataShapiro}b illustrate for $f=\SI{5.64}{\giga\hertz}$ the observation of all steps (odd and even), regardless of gate voltage $V_g$, or excitation frequency $f$. The response is similar to that of other systems such as carbon nanotubes \cite{Cleuziou2007}, graphene \cite{Heersche2007}, $\rm Bi_2Se_3$ \cite{Galletti2014} weak links, or the well-defined and meticulously analyzed case of atomic contacts \cite{Chauvin2006}. The amplitude of steps (along the bin counts axis) and height (along the voltage axis) are both reduced with decreasing frequency, and a correct resolution of the steps is only possible down to circa \SI{500}{\mega\hertz}. The evolution with power is shown in panel g). At zero microwave power ($P_{\rm RF}=0$), a single peak at $V=0$ indicates the supercurrent. As $P_{\rm RF}$ increases, Shapiro steps appear, starting from low values of $n$, while the amplitude of the supercurrent ($n=0$) decreases and eventually vanishes. For sufficiently high powers, an oscillatory pattern occurs in the amplitude of steps, as predicted for conventional Josephson junctions submitted to a voltage or current bias \cite{Tinkham2004,Russer1972}.

We also point out that we occasionally observe (in trivial and topological weak links) so-called subharmonic steps, i.e. steps for voltages $\frac{p}{q}\frac{hf}{2e}$ for $q=2$ or 3, and $p$ integer. These subharmonic steps are observed at high frequencies, in a regime where both conventional and topological weak links exhibit a conventional Josephson effect (see next paragraphs).
Such subharmonic steps are ubiquitous in Josephson junctions. They indicate a non-trivial phase locking condition, namely $\phi(t+q/f)=\phi(t)+2p\pi$. They are known to result from non-linearities, stray capacitive coupling between the superconducting electrodes, or higher harmonics in the current-phase relations. The latter have been predicted \cite{Tkachov2013} and detected \cite{Sochnikov2014} in our junctions. 

\subsubsection{Shapiro response of topological Josephson junctions}

As discussed in Section \ref{sec:1}, for topological Josephson junctions, the presence of $4\pi$-periodic supercurrents can in principle lead to the disappearance of odd steps, while even steps are preserved. First signs of the possible disappearance of the $n=1$ step have been reported in etched InAs nanowires \cite{Rokhinson2012}, driven by the predicted topological phase transition when a magnetic field is applied along the axis of the nanowire. Our observations made on the HgTe-based Josephson junctions \cite{Wiedenmann2016,Bocquillon2016} have conclusively improved the data, and exhibit the disappearance of several odd steps in devices made of 2D as well as 3D topological insulators.

\runinhead{Even sequence of Shapiro steps} Data obtained on 3D topological insulators are presented in the second column of Fig.\ref{fig:DataShapiro}. The $I$--$V$ clearly exhibits very strong $n=0 {\rm\, (supercurrent)}, 2, 4$ steps but  the steps $n=1$ and 3 are strongly suppressed. For higher voltages, the steps $n\geq5$ are not resolved at such power. In the original work of Wiedenmann {\it et al.}, only the $n=1$ step was missing \cite{Wiedenmann2016}. This new data thus shows a stronger $4\pi$-periodic behaviour and confirms that the disappearance of Shapiro steps is not related to hysteresis \cite{DeCecco2016}. Similar features have been observed in devices fabricated from HgTe-based topological insulators exhibiting the QSH effect \cite{Bocquillon2016}, and are summarized in the third column of Fig.\ref{fig:DataShapiro}. The line cut of panel c and the corresponding histogram (panel f) exhibits in particular the clear suppression of steps $n=1, 3, 5$. At even lower frequencies it has been possible to measure an even sequence of Shapiro steps up to $n=10$.

\runinhead{Dark fringes in the oscillatory pattern} The absence of odd steps is also remarkably clear on the colorplot of panel h of Fig.\ref{fig:DataShapiro} where the steps $n=1$ and 3 are suppressed, and $n=5$ weakly visible (3D TI), and in panel i for which the steps $n=1,3,5$ are suppressed (2D TI) for low microwave amplitude. Interestingly, the oscillatory pattern at higher microwave power is also modified : darker fringes (highlighted with white dashes) occur from the suppression of the first and third maxima of the oscillations. They suggest the progressive transformation from a $2\pi$- to a $4\pi$-periodic pattern with a halved period of oscillations. Despite some unexplained deviations, the colorplot of the 2D TI confirms the absence of odd steps and exhibits even more pronounced dark fringes in the oscillatory pattern. 

\begin{figure}[t]
\sidecaption[t]
\includegraphics[width=12 cm]{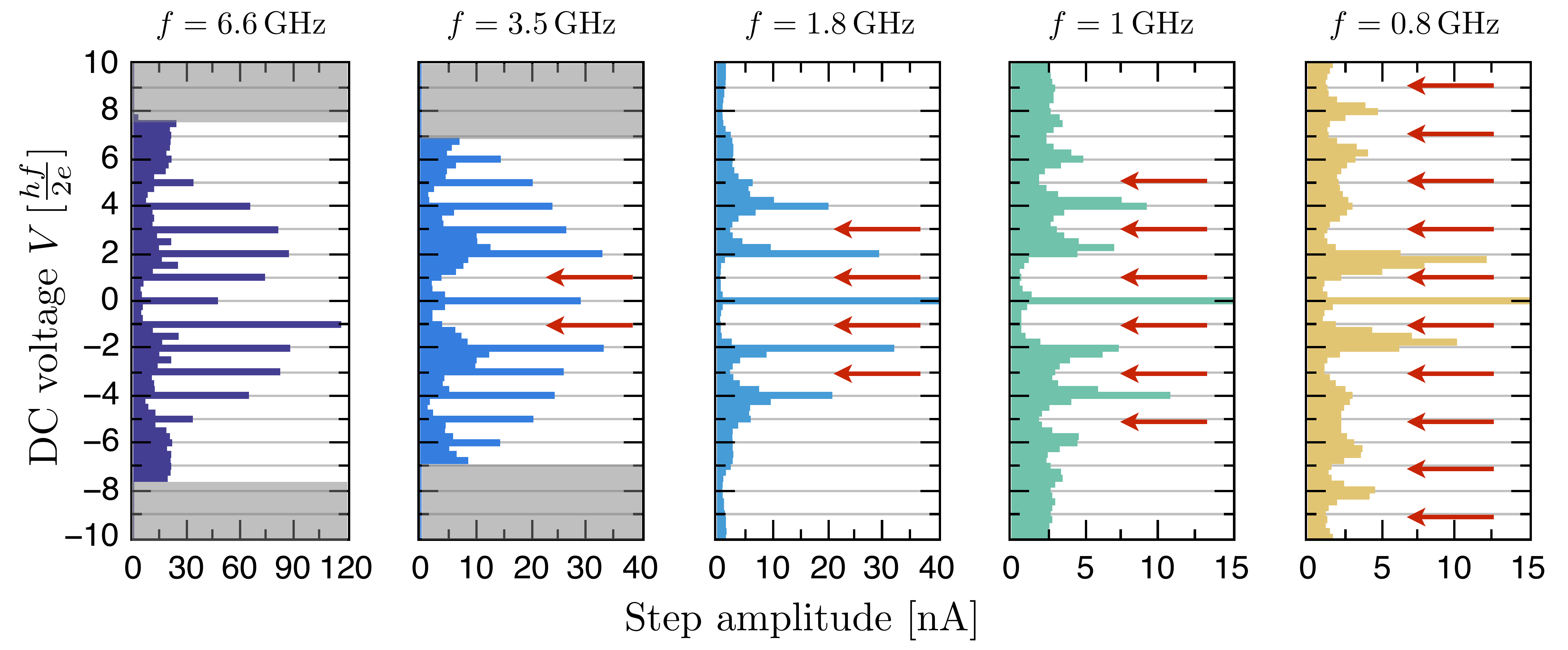}
\caption{Histograms of Shapiro steps -- Histograms of the voltage distribution obtained for different frequencies are shown. For a high frequency $f=\SI{6.6}{\giga\hertz}$, all steps are visibles. For lower frequencies, steps $n=1$ and 3 vanish ($f=\SI{1.8}{\giga\hertz}$), and up to $n=9$ at $f=\SI{0.8}{\giga\hertz}$. Missing odd steps are highlighted by red arrows.}
\label{fig:ShapiroFreqDep}
\end{figure}

\runinhead{Dependence on frequency} An important parameter that we emphasize now is the choice of the excitation frequency $f$. In analogy with the emission line at $f_{\rm J}/2$, only visible at low frequencies, the even sequence of Shapiro steps is only observed when $f$ is low. For high frequencies, the Shapiro response is conventional, and all step indices $n$ are present. As $f$ decreases, the odd steps progressively vanish, starting from low values of $n$. This is visible for example in Fig.\ref{fig:ShapiroFreqDep}. While odd steps are as visible as even ones at high frequencies ($f=\SI{6.6}{\giga\hertz}$), they are progressively suppressed as $f$ decreases. At the lowest accessible frequency ($f=\SI{0.8}{\giga\hertz}$), all odd steps up to $n=9$ are absent. This remarkable effect of frequency is similar to the one observed in the Josephson emission features, for which the line at $f_{\rm J}/2$ is mostly visible at low frequency. We analyze this behavior in Section \ref{subsec:RSJ}.

\section{Analysis : assessing the topological origin of the fractional Josephson effect}
\label{sec:TopoOrigin}

\begin{svgraybox}
The data summarized in this article exhibits two pieces of evidence of a strong $4\pi$-periodic {\it fractional Josephson effect}, despite the obstacles to its observation listed in Subsection \ref{subsec:obstacles}. We analyze in this section the possible (trivial or topological) origins of these features. First, we present an extended RSJ (Resistively Shunted Junction) model that includes a $4\pi$-periodic supercurrent. It enables a semi-quantitative analysis of our experimental results, and importantly yields an estimate of the amplitude of the $4\pi$-periodic supercurrent, compatible with a topological origin. Then, we analyze more in depth the effects of time-reversal and parity symmetry breaking, and Landau-Zener transitions, following the discussion of Section \ref{subsec:obstacles}.
\end{svgraybox}

\subsection{Modeling of a topological Josephson junction with $2\pi$- and $4\pi$-periodic modes}
\label{subsec:RSJ}
It is possible to calculate, at a microscopic level, the Andreev spectrum of a topological Josephson junctions in the zero bias equilibrium situation under various assumptions \cite{Fu2009,Olund2012,Tkachov2013,Tkachov2013a}. However, it is much more difficult to compute at the same elementary level the time-dependent response to a bias voltage or current, that emerges from the non-linear Josephson equations \cite{Sun2017,Li2017}, especially in the current-bias situation that is experimentally most relevant. To simulate the response of our devices, we turn to the RSJ model \cite{Stewart1968,McCumber1968,Russer1972}, in which we incorporate both $2\pi$- and $4\pi$-periodic supercurrents \cite{Dominguez2012,Wiedenmann2017,Bocquillon2016,Dominguez2017}.

\begin{figure}[t]
\sidecaption[t]
\includegraphics[width=11.9 cm]{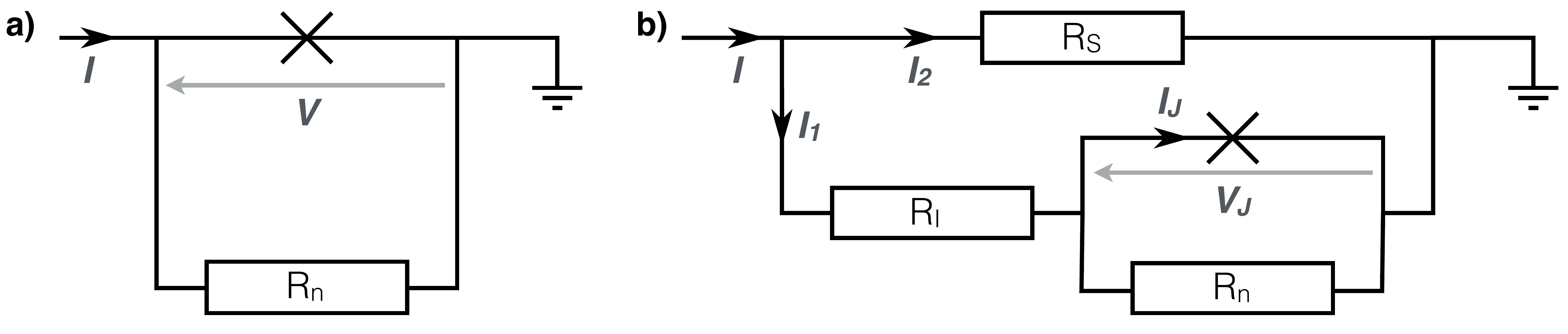}
\caption{Circuits for RSJ simulations -- a) Standard RSJ representation, with the Josephson junction carrying a supercurrent $I_S(\varphi)$ in parallel with a shunt resistance $R_n$. b) Modified RSJ circuit taking into account additional measurement setup with resistors $R_I$ and $R_S$. Adapted from \cite{Deacon2017}}
\label{fig:RSJcircuits}
\end{figure}

\runinhead{Framework of the RSJ model and modeling}
The RSJ model and its variants are commonly used to define the time-averaged voltage measured when a Josephson junction is submitted to a dc current bias. The universally valid time-evolution of the phase difference $d\phi/dt=2eV/\hbar$ is combined with the current-phase relation $I_{2D/3D}(\phi)$ derived from microscopic models (Section \ref{sec:FracJE}). The junction is associated to a resistive shunt to capture the ohmic transport of electrons (Fig.\ref{fig:RSJcircuits}a). Under current bias $I=I_{\rm dc}+I_{\rm rf}\sin (2\pi ft)$, one easily obtains a first order ordinary differential equation:
\begin{equation}
\frac{\hbar}{2e R_{n}}\dot\phi+I_s(\phi)=I_{\rm dc}+I_{\rm rf}\sin (2\pi ft)\label{eq:RSJ}
\end{equation}
For different values of $I_{2\pi}$ and $I_{4\pi}$, we simulate the results of this equation using a simple Runge-Kutta algorithm (RK4). We obtain the $I$--$V$ curve without ($I_{\rm rf}=0$) or with microwave excitation ($I_{\rm rf}\neq0$) to investigate Shapiro steps, or the Fourier spectrum of the voltage for the study of Josephson emission.

This model does not take into account all microscopic details, (for example $R_n$ and $I_s$ are assumed to be independent of voltage, which in reality may not be true). It nonetheless captures the key aspects of the dynamic Josephson current relevant to our observations. 
Besides, the more complicated bias circuit used to stabilize Josephson emission (Fig.\ref{fig:RSJcircuits}b) can be readily implemented. This circuit is indeed described by Eq.\ref{eq:RSJ}, with the substitutions $R_n\to \tilde R_n$ such that $\frac{1}{\tilde{R}_n}=\frac{1}{R_n}+\frac{1}{R_S}+\frac{R_I}{R_S R_n}$, and $I_c\to\tilde I_c = I_c  \big(1+\frac{R_I}{R_S}\big)$. Simulations performed in the standard RSJ model can be adapted to this new setup. The experimental data is then more naturally presented as a function of $I_1$ rather than $I$, which is obtained from $I_1=\frac{I-\frac{V_J}{R_S}}{1+\frac{R_I}{R_S}}$.

\subsubsection{Simulations of the response of a topological Josephson junction}

We give in this section a summary of the major results of the simulations. A comprehensive presentation of the simulation can be found in Refs.\cite{Wiedenmann2016,Deacon2017,Dominguez2017}.

\runinhead{Necessity of the $4\pi$-periodic contribution}

This framework enables a complete analysis of our results. As expected, the simulations emphasize the absolute necessity of a $4\pi$-periodic contribution to observe the vanishing of odd Shapiro steps or Josephson at half the Josephson frequency $f_{\rm J}/2$. Indeed, the exact definition of the $2\pi$-periodic supercurrent, namely the presence of higher harmonics ($\sin2\phi$ terms for example), have marginal influence on the observed disappearance of the odd steps, but the presence of a $4\pi$-periodic contribution is required to produce the signatures of the fractional Josephson effect.

This requirement contrasts with additional subharmonic Shapiro steps or Josephson emission at higher harmonics ($2f_{\rm J}$) which naturally appear in the simulations. They result from either non-sinusoidal current-phase relations or also from the addition of a small capacitive coupling in the RSJ equations (see below) \cite{Renne1974,Valizadeh2008}.

\begin{figure}[t]
\sidecaption[t]
\includegraphics[width=12 cm]{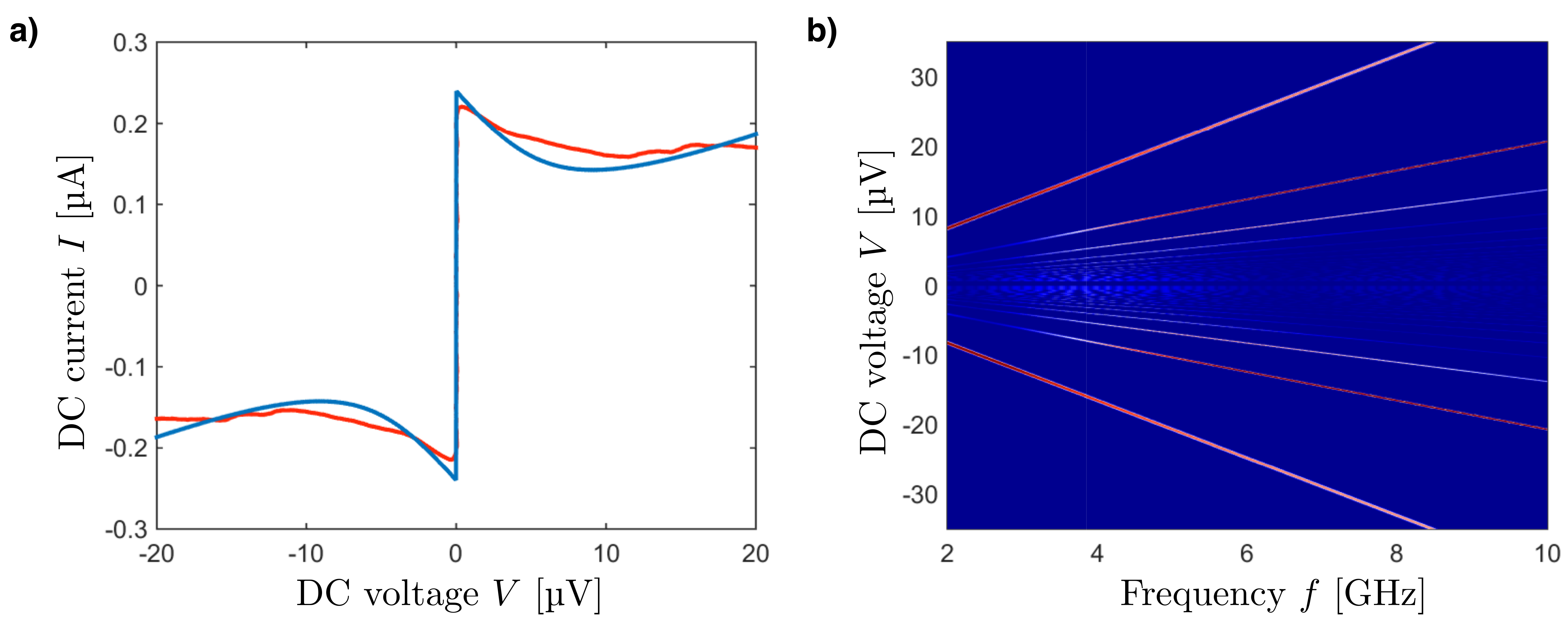}
\caption{Modeling Josephson Emission -- a) Simulated $I$--$V$ curve (blue) fitting measured data (red) The simulations are performed for $R_I=R_S=\SI{24}{\ohm}$ (nominal value of the resistors), $R_n=\SI{130}{\ohm},\, I_{4\pi}=\SI{100}{\nano\ampere}$, and $I_c=\SI{240}{\nano\ampere}$. b) Simulated Fourier transform of the voltage $V$ in the junction, as function of frequency $f$ and voltage $V$, with the same parameters as in a). Adapted from \cite{Deacon2017}}
\label{fig:RSJEmissionFit}
\end{figure}

\runinhead{Josephson emission spectrum and Shapiro steps} The numerical solutions to the RSJ equations give access to the time-dependent quantity $\phi(t)$, and allow for the study of the Fourier spectrum of the voltage, which is, (up to an unknown and frequency-dependent coupling factor) the quantity $A$ plotted in Section \ref{sec:emission}. First setting the critical current $I_c$ and the normal state resistance $R_n$ to optimally fit the experimental $I$--$V$ curve (see Fig.\ref{fig:RSJEmissionFit}a), we adjust the ratio $I_{4\pi}/I_{2\pi}$ (keeping $I_c$ constant) to obtain the best visual agreement between the experimental (Fig.\ref{fig:DataEmission}f) and simulated (Fig.\ref{fig:RSJEmissionFit}b) emission spectra.

In the present case, the simulated emission features reproduce semi-quantitatively the observed ones. In particular, as in the measurements, the $f_{\rm J}/2$ emission line dominates at low frequency while the $f_{\rm J}$ takes over at higher frequencies. However, the crossover between the two regimes at $f\simeq\SI{6}{\giga\hertz}$ is however smoother than the measured one.

Similarly, we have been able to reproduce sequences of even Shapiro using the extended RSJ model. In particular, we observe the vanishing of all Shapiro steps only when a $4\pi$-periodic supercurrent is present (see \cite{Wiedenmann2016}). As the excitation frequency $f$ decreases, we observe a transition from a conventional $2\pi$-periodic Shapiro step pattern to a fractional $4\pi$-periodic one. The crossover qualitatively describes our experimental observation, but in this model all odd steps vanish simultaneously, while our experiments exhibit a progressive disappearance starting from low values of the step index $n$ (see Fig.\ref{fig:ShapiroFreqDep}). We discuss the peculiar effect of frequency in the next paragraph.

\runinhead{Role of frequency - Estimating the $4\pi$-periodic supercurrent}
The puzzling dependence of the $f_{\rm J}/2$ emission line or the sequence of Shapiro steps in fact signals the transition from $2\pi$- to $4\pi$-periodic dynamics, and intrinsically results from the non-linearities of the RSJ equation (Eq.\ref{eq:RSJ}). It has first been elucidated and later clarified by Dominguez {\it et al.} \cite{Dominguez2012,Dominguez2017}, and we refer the reader to these two publications for details. 
The transition is controlled by the $4\pi$-periodic supercurrent $I_{4\pi}$ or equivalently the voltage $V_{4\pi}=R_nI_{4\pi}$ or the frequency scale $f_{4\pi}=\frac{2eR_nI_{4\pi}}{h}$. For an excitation frequency $f\ll f_{4\pi}$ (Shapiro steps) or a bias such that $V\ll V_{4\pi}$ (Josephson emission), the dynamics of the phase $\phi(t)$ is rather slow and very non-linear: $V(t)$ is a very anharmonic function of $t$, and becomes sensitive to the presence of the $4\pi$-periodic component. There, signatures of the fractional Josephson effect are very prominent. On the opposite, for $f\gg f_{4\pi}$ or $V\gg V_{4\pi}$, $V(t)$ is sinusoidal and rather insensitive to the $4\pi$-periodic contribution, so that the response of the device is comparable to conventional Josephson junction.

The identification of the crossover frequency in both experiments consequently provides a criterion to estimate the amplitude $I_{4\pi}$ of the $4\pi$-periodic supercurrent, and compare it to theoretical expectations.
We expect two modes to contribute in the 2D topological insulator (one edge mode on either edge), and one mode ($\theta=0$) for the 3D topological insulator. A perfectly transmitted mode carries in the short junction limit a supercurrent of maximum amplitude $\frac{e\Delta_{\rm i}}{h}$. In all tested devices, we found amplitudes ranging from \num{50} to \SI{300}{\nano\ampere}. Knowing $\Delta_{\rm i}$, this corresponds to 1 to 5 modes. This is roughly compatible with theoretical predictions, though it slightly exceeds them. It is likely that the large uncertainty on $\Delta_{\rm i}$ as well as the crude approximations of the RSJ model explain an improper estimate of $I_{4\pi}$. For example, we discuss below the influence of a capacitive coupling added to the RSJ equation (RCSJ model).

\subsubsection{Beyond the RSJ model}

Our analysis has so far been based on an extended RSJ model taking into account a $4\pi$-periodic contribution to the supercurrent. It provides a simple analysis and interpretation of our results, but several ingredients can be improved. 
The influence of a capacitive term in the RSJ equation can be important at microwave frequencies, and can be easily accounted for in the RCSJ model (Resistively and Capacitively Shunted Junction). This has been investigated in a recent study \cite{Pico2017}, which demonstrates that $I_{4\pi}$ can be overestimated with the above criterion even for small capacitances. Though it is difficult to estimate quantitatively $C$, it may explain the discrepancies on $I_{4\pi}$ between theory and experiments. Besides, in that model, the odd Shapiro steps are observed to vanish one by one (starting from low values of the step index $n$) rather than altogether in the RSJ model. This appears to be more in agreement with our data.
Based on our estimates \cite{Oostinga2013,Bocquillon2016}, the geometrical contribution to the capacitance between the two superconducting electrodes is however quite small. It has nevertheless be recently pointed out that Andreev bound states with high transparencies contribute to an intrinsic capacitance in mesoscopic devices, and may for example explain the observed hysteresis in the DC $I$--$V$ curves \cite{Antonenko2015}.

At this point, we would like to point out that it is easy to expand arbitrarily the non-linear differential equation with additional terms. Instead, we believe that the most rigorous but also most challenging approach is to construct a full microscopic understanding of the dynamics of the Andreev levels, including relaxation processes, and hence the dynamics of the currents and voltages.

\subsection{Time-reversal and parity symmetry breaking, and Landau-Zener transitions}

In this section, we review the influence of several mechanisms which are expected to obscure the fractional Josephson effect (Section \ref{subsec:obstacles}) but can not be easily taken into account in the preceding RSJ model.

\subsubsection{Time-reversal symmetry breaking} 

We have first pointed out that time-reversal symmetry in our devices imposes a degeneracy of Andreev bound states at $\phi\equiv0\bmod2\pi]$, which tends to restore the $2\pi$-periodicity due to enhanced parity relaxation, or on the opposite create $8\pi$-periodicity under the effect of electron-electron interactions. We have not detected any signal indicating an $8\pi$-periodic Josephson effect (such as Josephson emission at $f_{\rm J}/4$), but in contrast clearly observe a $4\pi$-periodic fractional Josephson effect. 

In our view, current models of topological Josephson junctions are partially inadequate and overlook important microscopic details. Though in our experiments time-reversal symmetry is not explicitly broken by a magnetic field or magnetic impurities, other mechanisms implicitly break time-reversal symmetry \cite{Strom2010,Crepin2012,Geissler2014,Vayrynen2014,Essert2015,Peng2016,Wang2017}, thus decoupling the Andreev spectrum from the continuum. Such mechanisms are for example in line with the observed weak stability of conductance quantization in the QSH regime \cite{Konig2007,Roth2009}. The role of two-particle inelastic scattering has also been recently highlighted \cite{Sticlet2018}.
Descriptions based on hard superconducting gaps also overlook the complexity of the density of states in induced systems \cite{Kopnin2011,Kopnin2014}, in which the behaviour of Andreev bound states remains poorly understood.

More experiments are required to characterize the induced superconducting state, with techniques such as point contact Andreev spectroscopy, already successfully employed on 3D topological insulators \cite{Blonder1982, Stehno2017,Wiedenmann2017}.

\subsubsection{Landau-Zener transitions and parity relaxation mechanisms}

The role of Landau-Zener transitions and parity relaxation mechanisms must be emphasized as they are expected to strongly influence the response of Josephson junctions, in particular at low frequency. These two processes cannot, to our understanding, easily be disentangled. We subsequently evaluate in parallel both possibilities.

\runinhead{Low-frequency behavior}
First, non-adiabatic Landau-Zener transitions can enforce a fractional Josephson effect from (sufficiently) driven gapped states. As a result of the Josephson equation $\frac{d\phi}{dt}=\frac{2eV}{\hbar}$, the phase $\phi(t)$ will vary faster as the voltage (or current) bias increases. Second, parity relaxation mechanism define a lifetime $\tau$ of the gapless Andreev bound states. When driven sufficiently rapidly, i.e. at characteristic frequencies $f$ such that $f\tau\gg1$, a topological Josephson junction can exhibit a $4\pi$-periodic response. On long times, the system however thermalizes to a conventional $2\pi$-periodic response.
In both cases, the Josephson effect should be conventional at low bias/low frequency, and turn into a fractional one above a crossover voltage $V_{c}$, as non-adiabatic processes are progressively activated. 

Our observations show in 2D topological insulators that the two signatures of a fractional Josephson effect are observed down to the lowest observables frequencies ($\simeq\SI{1}{\giga\hertz}$). Potential Landau-Zener transitions would thus be activated at a voltage $V_{LZ}\ll\SI{4}{\micro\volt}$. This sets an strict upper bound on a possible residual avoided crossing $\delta\ll\sqrt{\frac{V_{LZ}\Delta_{\rm i}}{8\pi}}=\SI{4}{\micro\electronvolt}$ \cite{Pikulin2012,Virtanen2013}, equivalent to large transmissions $D \gg 0.995$. This strong constraint on the transmission thus tends to exclude Landau-Zener transitions as the origin for the $4\pi$-periodic emission and suggests that the contributing bound states are indeed gapless.

Josephson junctions based on 3D topological insulators have mostly shown the same behavior (see \cite{Deacon2017}), with a strong $f_{\rm J}/2$ emission line at low frequency. The device presented in this article (Fig.\ref{fig:DataEmission}d) shows however a different behavior: a resurgence of the conventional $f_{\rm J}$ emission line is clearly visible for $f<\SI{2}{\giga\hertz}$ while the signal at $f_{\rm J}/2$ vanishes. Assuming Landau-Zener transitions are responsible for this transition around $V_{LZ}\simeq\SI{8}{\micro\volt}$, we find a gap on the order of $\delta=\SI{6}{\micro\volt}$ and a transmission $D\simeq 0.995$ (with $\Delta_{\rm i}=\SI{100}{\micro\volt}$).

\runinhead{Emission linewidths}

The finite lifetime of gapless Andreev bound states due to parity relaxation processes or Landau-Zener transitions can strongly affect the linewidth of the Josephson emission. For conventional Josephson radiation, the linewidth is in principle related to fluctuations in the pair or quasiparticle currents \cite{Stephen1968,Dahm1969} or can be dominated by the noise in the environment \cite{Likharev1986}. The linewidth at $f_{\rm J}/2$ can additionally reflect the influence of parity relaxation mechanisms \cite{Badiane2011,Badiane2013}. 

A difference in the linewidths of the two emission lines is clearly visible in our data. In the quantum wells, the extracted width ($\delta V_{2\pi}\simeq 0.5 - \SI{0.8}{\micro\volt}$) of the $f_{\rm J}$ emission line can be converted to a coherence time $\tau_{2\pi}=\frac{h}{2e\delta V_{2\pi}}\simeq 3 - \SI{4}{\nano\second}$. We find a shorter coherence time $\tau_{4\pi}\simeq 0.3-\SI{4}{\nano\second}$ for the $f_{\rm J}/2$ emission line. The linewidth is found to increase when the gate voltage $V_g$ is driven deeper in the conduction band. This may signal a decrease of lifetime when the $4\pi$-periodic modes are coupled to an increasing number of $2\pi$-periodic modes or to the continuum via ionization processes. We finally point out that the extracted lifetimes are consistent with the Shapiro steps being observable down to typically 0.5 to \SI{1}{\giga\hertz} \cite{Bocquillon2016}.

For the 3D topological device presented here, we find larger linewidths. The conventional $f_{\rm J}$ line with typically $\delta V_{2\pi}\simeq 2 - \SI{3}{\micro\volt}$ corresponding to $\tau_{2\pi}\simeq 0.7 - \SI{1}{\nano\second}$, while the fractional line has a width of $\delta V_{4\pi}\simeq 4 - \SI{7}{\micro\volt}$, yielding $\tau_{4\pi}\simeq 0.25 - \SI{0.5}{\nano\second}$. The difference of time scales between the 2D and 3D devices has been observed on several junctions but is not understood yet.

\section{Summary, conclusions and outlook}

We end this article with a summary of our observations, our conclusions and an outlook towards future experiments.

\runinhead{Summary and conclusions} The existence of $4\pi$-periodic supercurrents has been demonstrated from two sets of observations in Josephson junctions based HgTe in both a 2D and 3D topological insulator regime. First, these junctions show even sequences of Shapiro steps, with several missing odd steps (step indices $n=1,3,5$ missing). Secondly, they also exhibit strong Josephson emission at half the Josephson frequency $f_{\rm J}/2$ \cite{Deacon2017}. In contrast, reference devices made of graphene or trivial HgTe quantum wells do not show any of the above features, but a conventional Josephson response. Besides, we have presented a model for the junctions based on an extended RSJ model. It provides a simple yet efficient and plausible explanation of the transition from a $4\pi$- to a $2\pi$-periodic response with frequency or bias. This model yields the amplitude of the $4\pi$-periodic supercurrent, found to be compatible with theoretical predictions, though this estimate is subject to a large uncertainty. Our analysis of the emission features also sets very strong bounds on the possible influence of Landau-Zener activation, and provide additional information on the lifetime of the Andreev bound states. Finally, in junctions based on 2D topological insulators, the two signatures of the fractional Josephson effect are found to be concomitant (with respect to gate voltage). As presented in Fig.\ref{fig:JJCharac}b, they are more clearly visible when the current flow is mostly along the edges of the sample, and the bulk bands are depleted. The $4\pi$-periodic contribution is also detected in the whole $n$-conduction band. This suggests that the $4\pi$-periodic edge modes exist in parallel with bulk modes on the $n$-side. This interpretation is consistent with previous observations and predictions for HgTe \cite{Dai2008,Nowack2013,Hart2014}.

All in all, our observations thus strongly favor the presence of gapless Andreev bound states in our topological Josephson junctions, as initially predicted by Fu \& Kane \cite{Fu2008,Fu2009}. These devices, built from the well-characterized HgTe topological insulators, thus appear as first steps towards the development of a reliable platform for the future realization of Majorana end states and possibly scalable Majorana qu-bits.

\runinhead{Future objectives} The exact microscopic properties of the induced superconducting state, the presence of purely ballistic modes regardless of topology, or the role of time-reversal and parity symmetry breaking remains partially unclear. While our studies provide clear evidence of a supercurrent with $4\pi$-periodicity, a direct spectroscopy of such gapless Andreev bound states (ABS) is still missing. It is highly desirable as it would confirm the topological origin of the $4\pi$-periodicity and offer direct proof of the existence of gapless Majorana-Andreev bound states, as well as allow to verify the robustness of the topological protection.

This calls for a microscopic description of the induced superconductivity and of the dynamics of Josephson transport, including relaxation processes. Recent works have tackled this challenging program in topological systems \cite{Sun2017,Feng2018}, but the description of the induced superconductivity remains in many cases rudimental.

From an experimental point of view, some recent works have for example focused on point contact Andreev spectroscopy \cite{Blonder1982, Stehno2017,Wiedenmann2017} to probe the proximity effect in topological insulators in S-TI junctions. Future experiments will consequently focus on collecting more direct information on the Andreev spectrum forming in Josephson junctions, beyond their manifestation in the Josephson effect.

A first method consists in the study of the current-phase relation, which can be measured in asymmetric SQUIDs \cite{Zgirski2011,Delagrange2016}. By tuning the electron density via a gate, the goal is to identify the contribution of the topological modes. For example, the linear susceptibility $\frac{\partial I}{\partial \Phi}$ (with $\Phi$ the magnetic flux) at high frequency are very sensitive probes able to reveal the topologically-protected level crossing at $\phi=\pi$ \cite{Virtanen2013,Murani2017}. Furthermore, the investigation of the switching statistics \cite{Zgirski2011,Peng2016} around the critical current provides a means to prove that both states of a topological Andreev doublet have different parities.

The Andreev spectrum of a Josephson junction can also be obtained by means of tunneling spectroscopy, as already demonstrated in carbon nanotubes \cite{Pillet2010} or graphene \cite{Bretheau2017}. A reliable tunnel barrier could for example be obtained from hexagonal boron nitride flakes. The junction is then controlled via a phase bias mode by including the junction in a SQUID geometry, such that the phase difference $\phi$ across the junction is directly set by the magnetic flux through the ring.

A third method relies on microwave spectroscopy techniques: the absorption or emission of microwaves when at resonance with a transition in the Andreev spectrum is monitored \cite{Astafiev2010,Bretheau2013,VanWoerkom2017}. It can be performed in a SQUID geometry to bias the phase, with the SQUID inductively coupled to a microwave transmission line. Passing a microwave signal through the waveguide yields absorption lines of the Andreev spectrum of the junction. Additionally, the emission lines of the junction can also be measured.
Topological Majorana bound states (MBS) then show various characteristic features \cite{Peng2016}. First, since parity is invariable under photon absorption or emission, the transition between both states of the topological doublet should be strongly suppressed and only transitions involving the continuum should be visible. Second, the dispersion of the absorption/emission lines (as function of the magnetic flux) reflects the special $4\pi$-periodicity of the bound states. A natural follow-up to these experiments is then the exploration of topological transmons \cite{vanHeck2014} as a step towards braiding of Majorana Qubits.

\begin{acknowledgement}
We warmly thank the editors for their work and the opportunity to share our results. This work is supported by the German Research Foundation (Leibniz Program, SFB1170 Tocotronics) and the Elitenetzwerk Bayern program €œTopologische Isolatoren€, and the EU ERC-AG Program (project 4-TOPS). EB acknowledges support from the Alexander von Humboldt foundation. TMK acknowledges support from RSF Grant Non 17-72-30036 of the Russian Federation and Advanced Research Grant of the EC No. 339306 (METIQUM). RSD gratefully acknowledges support from "Grants-in-Aid for scientific research" (No. 16H02204), from the Japan Society for the Promotion of Science.
\end{acknowledgement}

\bibliographystyle{unsrt}
\bibliography{BibTMS2017.bib}
\end{document}